\documentclass[twocolumn,aps,prb,superscriptaddress]{revtex4-2}
\usepackage{color}
\usepackage{amsmath,amssymb}
\usepackage{pifont}
\usepackage{amssymb}  
\usepackage{bbold}
\usepackage{float}
\usepackage[normalem]{ulem}
\usepackage[utf8]{inputenc}
\usepackage{textgreek}
\usepackage{csquotes}
\usepackage[english]{babel}
\usepackage[labelfont=bf]{caption} 
\usepackage{tikz}
\usetikzlibrary{shapes.geometric, arrows.meta, positioning, shadows, backgrounds}
\usepackage{bbold}
\usepackage{makecell}
\usepackage{pifont}   
\usepackage{dcolumn}  
\usepackage{bm}       
\usepackage{multirow} 
\usepackage{placeins}
\usepackage[colorlinks]{hyperref}
\usepackage{mathtools}
\usepackage{subfigure}
\usepackage{amsmath}  
\usepackage{tikz}
\usetikzlibrary{quantikz2}

\usepackage{adjustbox} 
\usepackage[normalem]{ulem}
\usepackage{graphicx} 

\usepackage{breqn}
\usepackage{mathrsfs}

\usepackage{multirow}
\usepackage[margin=1in]{geometry}
\usepackage{subcaption}
\usepackage[ruled,vlined]{algorithm2e}
\usepackage{amsmath, amssymb}

\begin{document}

\title{A Variational Framework for Time-Dependent Quantum Systems with Applications to Floquet Hamiltonians}
\author{Ibsal Assi}
\affiliation{Department of Physics and Physical Oceanography$,$
Memorial University of Newfoundland and Labrador$,$ St. John’s$,$ Newfoundland $\&$ Labrador$,$ Canada A1B 3X7}

\author{Meenu Kumari}
\affiliation{Digital Technologies$,$ National Research Council Canada}
\affiliation{Perimeter Institute for Theoretical Physics$,$ Waterloo ON N2L 2Y5$,$ Canada}
\affiliation{Institute for Quantum Computing$,$ University of Waterloo$,$ Ontario N2L 3G1$,$ Canada}

\author{J. P. F. LeBlanc}
\affiliation{Department of Physics and Physical Oceanography$,$
Memorial University of Newfoundland and Labrador$,$ St. John’s$,$ Newfoundland $\&$ Labrador$,$ Canada A1B 3X7}
\affiliation{Compute Everything Technologies Ltd.$,$ St. John's$,$ Newfoundland $\&$ Labrador$,$ Canada}
\date{\today}

\begin{abstract}
We introduce a variational framework for approximating the time-evolution operator $\hat{U}(t)$ within a physically motivated operator manifold, reformulating quantum dynamics as a tractable problem in operator space using stationary action principle. For periodically driven systems, the resulting approximate evolution operator directly yields an effective Floquet Hamiltonian, offering a non-perturbative alternative to conventional expansion-based methods. The framework is systematically improvable by enlarging the operator pool and naturally incorporates symmetries and physical constraints. When the operator manifold is chosen from the terms of a truncated Magnus expansion, the variational procedure effectively resums the Magnus series within the restricted space, significantly enhancing accuracy. We benchmark the approach on the driven Rabi model, the driven Lipkin–Meshkov–Glick model, and the one-dimensional driven Ising chain, yielding effective Floquet Hamiltonians that are systematically more accurate than low-order Magnus expansions, particularly in regimes where the latter converge poorly, and illustrating applicability to systems with exponentially large Hilbert spaces. Although we focus here on Floquet systems, the formalism applies equally to generic time-dependent Hamiltonians, providing a versatile tool for non-equilibrium quantum dynamics.
\end{abstract}

\maketitle

\section{Introduction}
Time-dependent Hamiltonians emerge in a wide class of quantum systems, including adiabatic control \cite{PhysRevX.11.031070,PhysRevA.105.022618,Montenegro_Ferreira2026-it,n5xj-phwv}, quantum quenches \cite{PhysRevLett.110.257203,Caux2016-sm,PhysRevE.85.050102,Pelissetto2026-up}, time-dependent transport in mesoscopic systems \cite{PhysRevB.48.8487,PhysRevB.50.5528}, driven quantum impurities \cite{Deng2026-zy,Goker2008-vg}, time-dependent density functional theory \cite{Potthoff2011-fu,PhysRevA.95.042505}, and quantum systems under the influence of stochastic noise \cite{Seifi2024-yh}. 

Among these, a particularly important class of time-dependent Hamiltonians consists of those that are periodic in time, satisfying $\hat H(t+T)=\hat H(t)$ with period $T$. Such Hamiltonians arise naturally in a variety of settings, for instance a spin in an oscillating magnetic field, or electrons in a two-dimensional material illuminated by circularly polarized light \cite{Bukov2015-zb,Floquet_topological_insulators_1,Floquet_topological_insulators_2}. These systems are described by Floquet theory, which provides a powerful framework for analyzing their dynamics. Floquet's theorem states that the time-evolution operator can be decomposed as $\hat{U}(t)=\hat{P}(t)e^{-i\hat{H}_F t}$, where $\hat{P}(t)=e^{-i\hat{K}(t)}$ is the time-periodic micromotion operator (with $\hat{P}(nT)=I$ for integer $n$), $\hat{K}(t)$ is its Hermitian generator (often referred to as the kick operator), and $\hat H_F$ is the time-independent effective Hamiltonian, known as the Floquet Hamiltonian \cite{PhysRev.138.B979,PhysRevA.7.2203,Moessner2017-ey}.

Obtaining accurate effective Floquet Hamiltonians remains a central theoretical challenge in driven quantum systems and lies at the heart of Floquet engineering, where the goal is to design the time dependence of $\hat{H}(t)$ to realize desired properties of $\hat{H}_F$ \cite{Zhou2024-ld,Weitenberg2021-aw,PhysRevLett.123.090602,PhysRevLett.120.127601}.

Floquet engineering has emerged as a versatile tool across a wide range of experimental platforms, including ultracold atoms \cite{RevModPhys.89.011004,Weitenberg2021-aw,Zenesini2009,Jotzu2014-jn,Bakr2011-ez}, superconducting qubits \cite{SCqubits_review2020,NguyenPhysRevX2019,MatthewPRXQ2024,LauwensPRB2026}, and driven two-dimensional materials such as bilayer graphene \cite{2D_floq_1,2D_floq_2,TMLG_flat_bands_1,TMLG_flat_bands_2,TMLG_topology_1,TMLG_topology_2}. In these systems, periodic driving has enabled the realization of topological phases, controllable phase transitions, and noise-resilient quantum devices. Thus, reliable methods for computing $\hat{H}_F$ are not only of theoretical interest but also essential for the design and interpretation of experiments. 

From a theoretical perspective, access to the Floquet Hamiltonian greatly facilitates understanding of the exotic phenomena in driven systems. Consequently, various perturbative and non‑perturbative methods have been developed to obtain approximate Floquet Hamiltonians \cite{effective_HF_1,effective_HF_2,effective_HF_3,effective_HF_4,effective_HF_5}. Existing perturbative approaches, such as the Magnus expansion \cite{effective_HF_1}, provide analytical insight but are restricted to high-frequency regimes where the series converges. They also become increasingly cumbersome at higher orders due to nested commutators and multi-dimensional time integrals. Non-perturbative techniques, such as the flow equation approach \cite{Michael_Flow_eq}, have proven effective for generating accurate Floquet Hamiltonians across a wider range of driving parameters. However, this method typically requires an iterative, problem-specific construction of the flow Hamiltonian ansatz, which may need several refinement cycles. Moreover, as the operator manifold grows, repeated commutator evaluation becomes cumbersome and potentially infeasible for large Lie algebras. Exact numerical methods, on the other hand, become expensive for large Hilbert spaces and typically produce dense matrix representations rather than compact effective Hamiltonians expressed in terms of physically meaningful operators. A method that remains non-perturbative, systematically improvable, and capable of producing compact effective Floquet Hamiltonians is therefore highly desirable. Variational formulations provide a systematic framework for reducing complex dynamical problems to the optimization of a small set of parameters while preserving essential symmetries and conservation laws \cite{Frenkel1950,Dirac1930,McLachlan1964}.

In this work, we address this gap using a variational formulation of quantum dynamics based on an action principle for the time-evolution operator (henceforth also referred to as the propagator) \cite{Vogl2025-vk,Assi2026}. Specifically, we introduce a parameterized ansatz for the time-evolution operator $\hat U_A(\boldsymbol{\theta}(t))$ defined on a physically motivated operator manifold, where $\boldsymbol{\theta}(t)$ denotes a set of time-dependent variational parameters. Substituting this ansatz into the action principle yields a set of coupled equations of motion for the variational parameters, whose solution provides a variational approximation to the time-evolution operator over one driving period. From this, the effective Floquet Hamiltonian can be extracted directly. 

This contrasts with perturbative approaches which approximate the Hamiltonian; instead, we approximate the propagator directly, providing a non-perturbative framework that can be systematically improved through enlargement of the operator manifold. Unlike high-frequency expansions, our method is not restricted to weak driving or large frequencies. It naturally accommodates symmetries and physical constraints, and recasts the construction of effective Floquet Hamiltonians as a tractable variational problem in operator space.

Although Floquet systems are the primary focus of this work, the underlying variational principle is formulated at the level of the time-evolution operator and therefore applies equally well to non-periodic Hamiltonians. This generality is particularly relevant for quantum dynamics simulations, where accurate and efficient propagation of the time-evolution operator is a central challenge across many areas of physics, chemistry, and quantum information. The variational approach extends naturally to non-periodic drives, adiabatic protocols, and, with suitable modifications, to open quantum systems described by master equations. 

The remainder of this paper is organized as follows. In Sec. \ref{sec:theory}, we present our variational method (including the derivation of the equations of motion), discuss the Magnus expansion, introduce a variational resummation of the Magnus expansion, and review the numerical diagonalization in Sambe space. Sec. \ref{sec:applications} applies the method to three benchmark models: the Rabi model, the driven Lipkin–Meshkov–Glick model, and the driven Ising chain, comparing our results with exact diagonalization and other approximations. Finally, Sec. \ref{sec:conclusions} summarizes our findings and outlines future directions.

\section{Theory and methods}
\label{sec:theory}
This section outlines the theoretical framework and numerical techniques employed in this work. We first introduce our non-perturbative variational principle for the dynamics of time-dependent quantum systems. We then provide a concise overview of the Magnus expansion, which serves as the standard high-frequency approximation for comparison. Next, we demonstrate that our variational framework can be interpreted as a systematic resummation of the truncated Magnus series within a chosen operator manifold. Finally, we discuss numerical diagonalization in Sambe space, which provides numerically exact benchmark results whenever the Fourier-mode truncation is converged.

\subsection{Time-dependent variational principle for the evolution operator}
In this section, we present a non-perturbative variational approach based on an action principle for the time-evolution operator \cite{Vogl2025-vk,Assi2026}. We begin by introducing the action principle and deriving the equations of motion for the variational coefficients of a propagator ansatz. We then show how the Floquet Hamiltonian is extracted from the variational solution. Next, we discuss the scalability of the method to large Hilbert spaces using an operator manifold projection. Finally, we provide a rigorous error analysis that establishes a computable bound on the global error in terms of a local error rate. 

\subsubsection{The stationary action principle}
For a given time-dependent periodic Hamiltonian $\hat{H}(t)$ with period $T$, i.e. $\hat{H}(t+T)=\hat{H}(t)$, the time-evolution operator is described via Schr\"odinger equation
\begin{equation}
\label{eqn:SE_U}
    i\hbar\frac{\partial \hat{U}(t)}{\partial t}=\hat{H}(t)\hat{U}(t),
\end{equation}
where $\hbar$ is the (reduced) Planck's constant which we will set to unity throughout this work. The solution of the above equation is
\begin{equation}
    \hat{U}(t,0):=\hat{U}(t)=\hat{\mathcal{T}} {\rm exp}\Big\{-i\int_{0}^tdt' \hat{H}(t')\Big\}.
\end{equation}
where $\hat{\mathcal{T}}$ is the time-ordering operator. In general, $\hat{H}(t)$ and $\hat{H}(t')$ do not commute when $t\neq t'$, so $\hat{U}(t)$ cannot be evaluated in closed form. However, approximations exist such as the Magnus expansion \cite{effective_HF_1}, Krylov subspace methods \cite{PRL_Krylov_Methods}, and other time-dependent variational principles \cite{RevModPhys.44.602}.

We start our variational formalism by introducing the following action \cite{Assi2026,Vogl2025-vk} 
\begin{equation}
\label{eqn:main_action}
    \mathcal{S}[\hat{U},\hat{U}^\dagger]=\int dt\, \mathrm{Tr}\left[\hat{U}^\dagger(t)\Big(i\frac{\partial \hat{U}(t)}{\partial t}-\hat{H}(t)\hat{U}(t)\Big)\right].
\end{equation}
Treating the entries of $\hat{U}(t)$ and $\hat{U}^\dagger(t)$ as independent variational parameters and imposing the stationary condition $\delta \mathcal{S}[\hat{U},\hat{U}^\dagger]=0$ yields the Schr\"odinger equation Eq.~\eqref{eqn:SE_U}. We note that this action is not unique; other functionals, such as the squared residual action \cite{Assi2026,Vogl2025-vk}, exist. The procedure described below is general and extends naturally to such alternatives.

In practice, one would choose a time-evolution ansatz $\hat{U}_A(\boldsymbol{\theta}(t))$ that is parameterized by a set of variational parameters $\{\theta_j(t)\}_{j=1}^{j=M}$. By replacing the exact unitary $\hat{U}(t)$ in Eq. \eqref{eqn:main_action} with $\hat{U}_A(\boldsymbol{\theta}(t))$ and extremizing the action with respect to the variational parameters, we obtain the following equations of motion (EOM) for the variational parameters
\begin{equation}
\label{eqn:ODE}
    \sum_k g_{jk}(\boldsymbol{\theta}(t))\dot{\theta}_k(t)=f_j(\boldsymbol{\theta}(t);t),
\end{equation}
where
\begin{equation}
\label{eqn:Qmetric}
    g_{jk}(\boldsymbol{\theta}(t))=\mathrm{Tr}\left[\Big(\frac{\partial \hat{U}_A(t)}{\partial \theta_j}\Big)^\dagger\frac{\partial \hat{U}_A(t)}{\partial \theta_k}\right],
\end{equation}
is the quantum geometric tensor (QGT) and
\begin{equation}
\label{eqn:GF}
    f_j(\boldsymbol{\theta}(t);t)=-i\mathrm{Tr}\left[\Big(\frac{\partial \hat{U}_A(t)}{\partial \theta_j}\Big)^\dagger\hat{H}(t)\hat{U}_A(t)\right],
\end{equation}
is the generalized force \cite{Assi2026}. The solution to the system of coupled differential equations in Eq. \eqref{eqn:ODE} can be obtained via different numerical techniques such as Runge-Kutta methods. We should note that in certain cases, $g$ can be a singular matrix and one can use the Pseudo-inverse when performing numerical integration of the EOM. 

Although this work is focused on time-periodic Hamiltonians, we stress that the above formulation is valid even if $\hat{H}(t)$ is \emph{not} periodic in time. This enables the application of our method to a wider class of problems beyond the examples studied in this work.

\subsubsection{The time-evolution operator ansatz}
\label{sec:moderate_N_procedure}
In this work, we consider the following time-evolution ansatz:
\begin{equation}
    \hat{U}_A(\boldsymbol{\theta}(t))=e^{-i\hat{A}(\boldsymbol{\theta}(t))}
\end{equation}
where $\hat{A}(\boldsymbol{\theta}(t))=\sum_j\theta_j(t)\hat{\mathcal{O}}_j$, with $\{\hat{\mathcal{O}}_j\}$ a pool of Hermitian operators and $\theta_j(t)$ are real variational parameters with initial conditions $\theta_j(0)=0$ for all $j$.

The choice of this operator pool is problem-dependent and guided by physical intuition: it may be based on the Lie algebra generated by the static Hamiltonian and drive, on the symmetries of the system, or on the operators appearing in a low-order Magnus expansion. For many-body systems, we typically include local and nearest-neighbor interactions, with the option to systematically enlarge the pool by adding higher-order (e.g., three-body) terms to improve accuracy.  

This ansatz applies to arbitrary time dependence: the Magnus expansion produces exactly this form once nested commutators are evaluated. For periodic systems, Floquet theory gives $\hat{U}(t)=e^{-i\hat{K}(t)}e^{-i\hat{H}_F t}$, and the Baker-Campbell-Hausdorff (BCH) formula reduces this product to a single exponential when the operator pool forms a closed Lie algebra. Thus, $\hat{A}(t)$ naturally captures both the micromotion and stroboscopic evolution, becoming exact whenever the pool is closed under commutation and contains the dynamical algebra generated by $\hat{H}(t)$.

The goal is to select a pool that allows the ansatz to approximate the exact dynamics over one period while its size $M$ remains significantly smaller than the full Hilbert space dimension. This balances computational efficiency with accuracy and ensures that the resulting effective Hamiltonian is compact and physically transparent. Once the variational evolution over one period is obtained, the approximate Floquet Hamiltonian follows directly as
\begin{equation}
    \hat{H}_F=\sum_j\frac{\theta_j(T)}{T}\hat{\mathcal{O}}_j.
\end{equation}
It remains to discuss the EOM and their solutions. First, we derive an expression for the QGT (Eq. \eqref{eqn:Qmetric}) associated with the above ansatz. As demonstrated in Appendix \ref{app:g_and_f_integrals} (Eq. \eqref{eq:single_integral_g}), we obtain
\begin{equation}
\label{eq:g_single_integral}
    g_{jk}=\int_{-1}^{+1}ds (1-|s|)\mathrm{Tr}\left[\hat{\mathcal{O}}_je^{is\hat{A}(\boldsymbol{\theta}(t))}\hat{\mathcal{O}}_ke^{-is\hat{A}(\boldsymbol{\theta}(t))}\right].
\end{equation}
which is real and symmetric since the basis operators are Hermitian and $\theta_j(t)$ are real. In moderate Hilbert space dimensions, a further simplification can be made. Let us assume that $\lambda_n(t)$ is an eigenvalue of $\hat{A}(\boldsymbol{\theta}(t))$ with $\ket{\phi_n(t)}$ being the corresponding eigenvector, the above expression becomes 
\begin{equation}
\label{eq:g_doubel_sum}
g_{jk}
=
\sum_{m,n}
{\rm sinc}^2\left(\Delta_{nm}(t)\right)
\left[\hat {\mathcal{O}}_j\right]_{nm}\left[\hat {\mathcal{O}}_k\right]_{mn} .
\end{equation}
where ${\rm sinc}(x)=\sin(x)/x$, $\Delta_{nm}(t)=\frac{1}{2}(\lambda_n(t)-\lambda_m(t))$ and $\left[\hat {\mathcal{O}}_j\right]_{nm}=\langle \phi_n(t)|\hat {\mathcal{O}}_j|\phi_m(t)\rangle$. Defining the Kernel matrix $\mathcal{K}^{g}_{mn}={\rm sinc}^2\left(\Delta_{mn}(t)\right)$, we can simply rewrite the QGT as
\begin{equation}
\label{eqn:gjk_final}
g_{jk}=\mathrm{Tr}\left[\hat{\mathcal{O}}_j\times (\mathcal{K}^{g}\circ \hat{\mathcal{O}}_k)\right]
\end{equation}
where $\circ$ denotes the Hadamard matrix product (element-wise product) and $\times$ is the usual matrix product.

The second ingredient of the EOM is the generalized force. As shown in Appendix \ref{app:g_and_f_integrals}, we find the following expression for $f_j$
\begin{equation}
\label{eq:f_doubel_sum}
    f_j=\int_0^1 ds\, \mathrm{Tr}\left[e^{-is\hat{A}(\boldsymbol{\theta}(t))}\hat{\mathcal{O}}_je^{is\hat{A}(\boldsymbol{\theta}(t))}\hat{H}(t)\right].
\end{equation}
representing a real quantity for Hermitian Hamiltonians. Following the same derivations outlined above, the expression can be further simplified to
\begin{equation}
\label{eqn:fj_final}
f_j
=\mathrm{Tr}\left[(\mathcal{K}^{f}\circ \hat{\mathcal{O}}_j)\times\hat H(t)\right],
\end{equation}
where $\mathcal{K}^{f}_{nm}=i\frac{e^{-2i\Delta_{nm}} - 1}{2\Delta_{nm}}$ is the Kernel of $f_j$.

\begin{figure*}[t]
    \centering
    \includegraphics[width=0.9\textwidth]{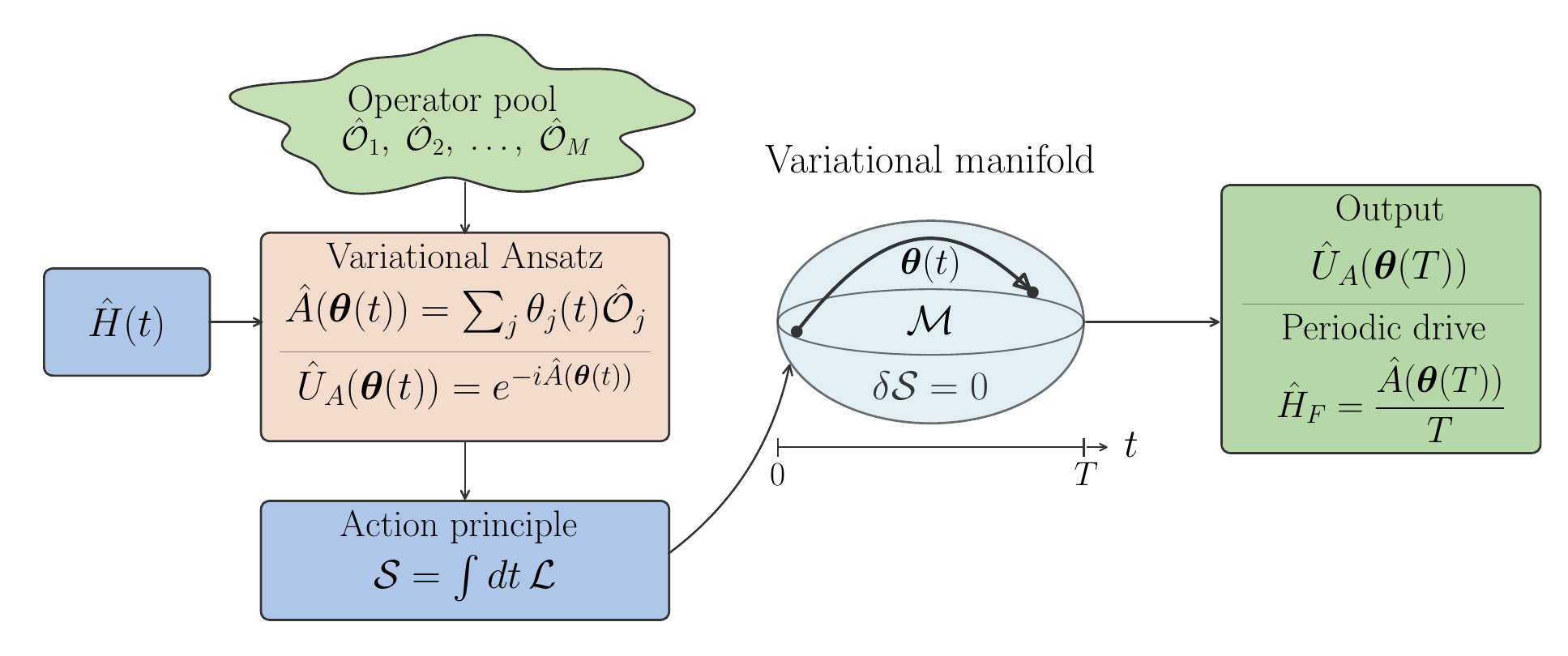}
    \caption{Schematic illustration of the variational framework. Given a time-periodic Hamiltonian $\hat{H}(t)$ and a physically motivated pool of operators $\{\hat{\mathcal O}_j\}$, we construct a variational ansatz $\hat{U}_A(\boldsymbol{\theta}(t))=e^{-i\hat{A}(\boldsymbol{\theta}(t))}$. The action $\mathcal{S}$ is optimized to obtain the equations of motion for the variational parameters $\boldsymbol{\theta}(t)$. After numerical integration yields the approximate propagator $\hat{U}_A(\boldsymbol{\theta}(T))=e^{-i\hat{A}(\boldsymbol{\theta}(T))}$. If the Hamiltonian is periodic, $\hat{H}(t+T)=\hat{H}(t)$, the Floquet Hamiltonian is extracted as $\hat{H}_F=\hat{A}(\boldsymbol{\theta}(T))/T$}
    \label{fig:scheme}
\end{figure*}

Having found those compact formulas for $g_{jk}$ and $f_j$, one can then use numerical integrations to solve Eq.\eqref{eqn:ODE} and obtain the variational parameters $\{\theta_j(t)\}$. Finally, we have included a schematic diagram summarizing our variational method as shown in Fig. \ref{fig:scheme}.  

\subsubsection{Scalability to larger system sizes}
\label{sec:scalability}
In the previous section, we provided simple expressions for $g_{jk}$ and $f_j$ that rely on diagonalizing $\hat{A}(\boldsymbol{\theta}(t))$. While mathematically correct, these expressions become impractical for large systems because the Hilbert space grows exponentially with the number of degrees of freedom. For example, in a spin-$1/2$ chain, $\hat{A}(\boldsymbol{\theta}(t))$ is a $2^N\times 2^N$ matrix; for large $N$, full diagonalization is infeasible. Moreover, this diagonalization would be required at every time step of the ODE integration, making it the main computational bottleneck. We therefore need an alternative approach to extend our variational method to large systems.

The key idea is that the number of variational parameters is set by the size $M$ of the pool of basis operators $\{\hat{\mathcal{O}}_j\}$, which is generally chosen to be \emph{much smaller} than the actual Hilbert space dimension ($\dim \mathcal{H}$). The exponentially
large matrices only ever enter through the traces in the definitions of $g_{jk}$ and
$f_j$. The strategy of this section is therefore to reformulate those traces so that the entire calculation lives in the $M$-dimensional
operator space and the $\dim \mathcal{H}\times \dim \mathcal{H}$ operators never appear explicitly. 

By projecting the dynamics onto this operator manifold, we can use the relation
\begin{equation}
[\hat{\mathcal{O}}_j,\hat{\mathcal{O}}_k]=i\sum_\ell \alpha_{jk}^\ell \hat{\mathcal{O}}_\ell
\end{equation}
where contributions from operators outside this pool are projected out. Here $\alpha_{jk}^\ell$ are real parameters that can be obtained and stored once for a fixed pool of operators. If the operators $\{\hat{\mathcal{O}}_j\}$ form an orthogonal basis, i.e. $\mathrm{Tr}\left[\hat{\mathcal{O}}_j\hat{\mathcal{O}}_k\right]\propto \delta_{jk}$, then evaluating the commutator directly yields 
\begin{equation}
\label{eqn:alphas}
\alpha_{jk}^\ell=\frac{\mathrm{Tr}\left[[\hat{\mathcal{O}}_j,\hat{\mathcal{O}}_k]\hat{\mathcal{O}}_\ell\right]}{\mathrm{Tr}\left[\hat{\mathcal{O}}_\ell^2\right]}
\end{equation}
For non-orthogonal basis, one can determine these parameters by minimizing $||[\hat{\mathcal{O}}_j,\hat{\mathcal{O}}_k]-i\sum_\ell \alpha_{jk}^\ell \hat{\mathcal{O}}_\ell||_F$. 

Returning to Eq. \eqref{eq:g_single_integral}, we define $\hat{\mathcal{O}}_k(s)=e^{is\hat{A}(\boldsymbol{\theta}(t))}\hat{\mathcal{O}}_ke^{-is\hat{A}(\boldsymbol{\theta}(t))}$. Differentiating with respect to $s$ gives
\begin{equation}
    \frac{d\hat{\mathcal{O}}_k(s)}{ds}=i[\hat{A}(\boldsymbol{\theta}(t)),\hat{\mathcal{O}}_k(s)].
\end{equation}

Projecting the dynamics onto the manifold $\{\hat{\mathcal{O}}_j\}$, we write $\hat{\mathcal{O}}_k(s)=\sum_{p}c_{p}^{(k)}(s)\hat{\mathcal{O}}_p$ with real coefficients $c_{p}^{(k)}(s)$. Using $\hat{A}(\boldsymbol{\theta}(t))=\sum_j\theta_j(t)\hat{\mathcal{O}}_j$ and the commutator expansion, we obtain the following EOM for $c_{p}^{(k)}(s)$
\begin{equation}
\label{eq:EOM_cs}
    \frac{dc_{\ell}^{(k)}(s)}{ds}=-\sum_{p}\chi_{\ell p} c_{p}^{(k)}(s),
\end{equation} 
where
\begin{equation}
\label{eqn:chi_ellp}
    \chi_{\ell p}=\sum_j \theta_j(t)\alpha_{jp}^\ell.
\end{equation}
Defining the column vector $c^{(k)}(s)=(c_{1}^{(k)}(s),\cdots,c_{M}^{(k)}(s))^t$, Eq.~\eqref{eq:EOM_cs} becomes
\begin{equation}
    \frac{dc^{(k)}(s)}{ds}=-\chi c^{(k)}(s).
\end{equation}
whose solution is $c^{(k)}(s)=e^{-s\chi}c^{(k)}(0)$. The initial vector $c^{(k)}(0)$ has a single non-zero entry, namely $+1$ at the kth position. Substituting this result into the expression of the QGT  \eqref{eq:g_single_integral} yields the matrix form
\begin{equation}
\label{eqn:g_mat_integral}
    g=\Phi \int _{-1}^{+1}ds\, (1-|s|)e^{-s\chi},
\end{equation}
where $\Phi$ is the overlap matrix defined as
\begin{equation}
\Phi_{jk}=\mathrm{Tr}\left[\hat{\mathcal{O}}_j\hat{\mathcal{O}}_k\right].
\end{equation}
For local basis operators, $\Phi$ can be computed analytically without numerical traces, as illustrated in Appendix \ref{app:Ising_relations} for our driven Ising model. To evaluate the integral in Eq. \eqref{eqn:g_mat_integral}, we perform an eigendecomposition of $\chi$: $\chi=V\Lambda V^{-1}$, where the columns of $V$ are the eigenvectors of $\chi$ and $\Lambda$ is a diagonal matrix of eigenvalues. Using the identity
\begin{equation}
    \int _{-1}^{+1}ds\, (1-|s|)e^{-s\chi}=2V\frac{\cosh(\Lambda)-1}{\Lambda^2}V^{-1},
\end{equation}
we obtain the closed-form expression
\begin{equation}
\label{eqn:g_mat_integral_final}
    g=2\Phi V\frac{\cosh(\Lambda)-1}{\Lambda^2}V^{-1}.
\end{equation} 

It thus remains to find a similar expression for the generalized force $f_j$ in Eq. \eqref{eqn:GF}. Writing the Hamiltonian as $\hat{H}(t)=\sum_k h_k(t)\hat{\mathcal{O}}_k$, we express $f_j$ as
\begin{equation}
f_j=\sum_kh_k(t)\int_0^1 ds\, \mathrm{Tr}\left[\hat{\mathcal{O}}_j\hat{\mathcal{O}}_k(s)\right].
\end{equation}
Using $\hat{\mathcal{O}}_k(s)=\sum_p c_p^{(k)}(s)\hat{\mathcal{O}}_p$ and $c^{(k)}(s)=e^{-s\chi}c^{(k)}(0)$, we obtain
\begin{equation}
    f_j=\sum_{k}h_k(t)\int_0^1 ds\, \left[\Phi e^{-s\chi}\right]_{jk},
\end{equation}
or, in matrix form,
\begin{equation}
\label{eqn:f_mat_integral_final}
    f=\Phi V \frac{1-e^{-\Lambda}}{\Lambda}V^{-1}\vec{h}(t),
\end{equation}
where $\vec{h}(t)$ is a column vector with components $h_k(t)$. It is important to note that both $g$ and $f$ have a common left factor $\Phi$. Provided $\Phi$ is invertible (which is the case of linearly independent basis, e.g. in multi-qubit Hamiltonians), this factor cancels out in the equations of motion $g\dot{\theta}=f$.  

The expressions above demonstrate that both the quantum metric $g_{jk}$ and the generalized force $f_j$ can be evaluated using only the small operator basis $\{\hat{\mathcal{O}}_j\}$ of size $M$, without ever constructing the full Hilbert space. All operations reduce to matrix algebra on $M\times M$ matrices, where $M$ is typically much smaller than the Hilbert space dimension. If the set $\{\hat{\mathcal{O}}_j\}$ is closed under commutation (i.e., it forms a Lie algebra), the derivation is exact and the dynamics remains strictly within the manifold spanned by these operators. In practice, the pool may be chosen as a physically motivated truncated set; the resulting dynamics then corresponds to a projection onto that manifold. The accuracy can be systematically improved by enlarging the pool until the closure condition is approximately satisfied, making this approach a powerful and scalable tool for simulating large quantum systems.

Alternatively, one can compute $g$ and $f$ numerically as follows. The integrals over $s$ in Eqs. \eqref{eq:g_single_integral} and \eqref{eq:f_doubel_sum} can be evaluated using standard numerical quadrature, while the traces are estimated using stochastic trace estimation \cite{Meyer2021-mt,Ubaru2017-wq}. The required action of the matrix exponentials $e^{\pm is\hat{A}}$ on these vectors can be computed using Trotter-Suzuki decompositions or Krylov subspace methods \cite{trotter1959product,casares2026theorypracticetrotterproduct,maxwell2026practicalestimationtrottererror,Saad2003}. This approach also avoids diagonalizing $\hat{A}$ and is therefore applicable to large many-body systems, at the expense of quadrature, statistical, and propagation errors. The interested reader can easily pursue these directions.

\subsubsection{Error analysis and rigorous error bounds}
\label{sec:err_analysis}
A central question in any variational approach to quantum dynamics is how to assess the accuracy of the approximation without prior knowledge of the exact solution. The global error $\eta(t)$, defined below as the normalized Frobenius distance between the exact and approximate propagators, provides a direct measure of fidelity. However, its evaluation requires access to the exact time-evolution operator, which is precisely the object we seek to approximate, rendering $\eta(t)$ itself inaccessible in practice for large systems. 

To overcome this limitation, we derive a computable upper bound on $\eta(t)$ in terms of a local error rate $\epsilon(t)$, which can be readily monitored during the simulation. This local quantity captures the instantaneous violation of the Schr\"odinger equation, and its time integral yields a rigorous yet practically computable upper bound on the accumulated global error. We term this bound the \emph{accumulated error bound} (AEB) and employ it throughout as a diagnostic tool for assessing the reliability of the variational approximation.

Let $\hat{U}(t,0)$ be the exact time-evolution operator for the time-dependent Hamiltonian $\hat{H}(t)$, satisfying $i\dot{\hat{U}}(t,0)=\hat{H}(t)\hat{U}(t,0)$ with $\hat{U}(0,0)=\mathbf{1}$, and let $\hat{U}_A(t)$ be the variational approximation to $\hat{U}(t,0)$. We define the global error as
\begin{equation}
\label{eqn:global_error}
    \eta(t)=\frac{\left\|\hat{U}(t,0)-\hat{U}_A(t)\right\|_F}{2\sqrt{D}},
\end{equation}
where $\left\|\cdot\right\|_F$ denotes the Frobenius norm $\left( \left\| X\right\|_F = \sqrt{\mathrm{Tr}\left(XX^{\dagger}\right)} \right)$ and $D$ is the Hilbert space dimension. The normalization by $2\sqrt{D}$ ensures that $0\leq\eta(t)\leq 1$, since the Frobenius norm of the difference between any two unitary operators is at most $2\sqrt{D}$.

A key quantity is the local residual 
\begin{equation}
\label{eqn:local_residual}
    \hat{R}(t)=i\hat{\dot{U}}_A-\hat{H}(t)\hat{U}_A
\end{equation}
which measures the extent to which the approximate propagator fails to satisfy the Schr\"odinger equation and vanishes identically for the exact dynamics. This residual directly determines the evolution of the error operator
\begin{equation}
    \hat{\Delta}(t)=\hat{U}(t,0)-\hat{U}_A(t)
\end{equation}
which encodes the accumulated deviation of the variational approximation from the exact propagator. Adding and subtracting $\hat{H}(t)\hat{U}(t,0)$ and $i\hat{\dot{U}}(t)$, respectively, from Eq.\eqref{eqn:local_residual} (which leaves $\hat{R}(t)$ unchanged), yields
\begin{equation}
\label{eqn:delta_ode}
    \hat{\dot{\Delta}}(t)=i\hat{R}(t)-i\hat{H}(t)\hat{\Delta}(t).
\end{equation}

This linear inhomogeneous operator differential equation is solved as follows. We make the ansatz $\hat{\Delta}(t)=\hat{U}(t,0)\hat{W}(t)$. Substituting into Eq. \eqref{eqn:delta_ode} and using the Schr\"odinger equation for $\hat{U}(t,0)$ gives
\begin{equation}
    \hat{U}(t,0)\hat{\dot{W}}(t)=i\hat{R}(t),
\end{equation}
and since $\hat{U}(0,t)$ is unitary, $\hat{U}^\dagger(t,0)=\hat{U}(0,t)$. Multiplying on the left by $\hat{U}^\dagger(t,0)$ and integrating from $0$ to $t$ with $\hat{W}(0)=0$ yields
\begin{equation}
    \hat{W}(t)=i\int_0^t ds\,\hat{U}(0,s)\hat{R}(s).
\end{equation}
Thus,
\begin{equation}
    \hat{\Delta}(t)=i\int_0^t ds\,\hat{U}(t,0)\hat{U}(0,s)\hat{R}(s)
\end{equation}
Using the composition property of the time-evolution operator $\hat{U}(t,0)\hat{U}(0,s)=\hat{U}(t,s)$, we obtain the final formula,
\begin{equation}
    \hat{\Delta}(t)=i\int_0^t ds\,\hat{U}(t,s)\hat{R}(s),
\end{equation}
Using the triangle inequality, we obtain
\begin{equation}
\label{eqn:triangular_inequality}
    \left\|\hat{\Delta}(t)\right\|_F\leq \int_0^t ds\,\left\|\hat{R}(s)\right\|_F
\end{equation}
where we used the fact that for a unitary $\hat{U}(t,s)$, the Frobenius norm $\left\|\hat{U}(t,s)\hat{R}(s) \right\|_F$ simplifies to $\left\|\hat{R}(s) \right\|_F$. Dividing Eq. \eqref{eqn:triangular_inequality} by $2\sqrt{D}$ and using Eq.\eqref{eqn:global_error}, we obtain the inequality
\begin{equation}
\label{eqn:central_inequality}
    \eta(t)\leq \int_0^t ds\, \epsilon(s)
\end{equation}
where
\begin{equation}
    \epsilon(t)=\frac{\left\|\hat{R}(t)\right\|_F}{2\sqrt{D}}
\end{equation}
is the local (residual) error rate. The integral $\int_0^t ds\, \epsilon(s)$ defines the accumulated error bound (AEB). This is the main result of this section: the global error is bounded by the time-integral of the local error rate, providing a rigorous upper bound on the accuracy of the approximation. In particular, maintaining a small local error rate throughout the simulation guarantees a strict upper bound on the global error. Although the AEB provides only an upper bound on the global error and therefore does not mathematically guarantee that a smaller AEB corresponds to a smaller $\eta(t)$. However, all of the examples considered in this work exhibit this correlation when comparing our variational method with the Magnus expansion.

As demonstrated in Appendix \ref{app:optimum_var_vs_mag}, the variational equations \eqref{eqn:ODE} are derived by minimizing the instantaneous squared residual $\Phi=\|\hat{R}\|_F^2/2$ with respect to the parameter velocities $\dot{\theta}_j$. Since the local error rate is given by $\epsilon(t)=\|\hat{R}(t)\|_F/(2\sqrt{D})$, this corresponds to minimizing $\epsilon(t)$ at each instant for the current variational parameters. Consequently, the variational trajectory is locally optimal in the sense that it minimizes the instantaneous contribution to the accumulated error bound in Eq. \eqref{eqn:central_inequality}. Thus, within the chosen operator manifold, the variational method produces the locally tightest error bound. This local optimality, however, does not imply that the resulting error rate $\epsilon(t)$ or the accumulated error bound is necessarily smaller than that obtained from a different approximation scheme, such as a truncated Magnus expansion, since such methods are not generated by the same local optimization principle and generally follow different approximate trajectories. Nevertheless, as demonstrated by all examples considered in this work, smaller accumulated error bounds consistently correlate with smaller global errors. 

It remains to show that the local error rate $\epsilon(t)$ can be computed efficiently without constructing the full Hilbert space. Using $\hat{\dot{U}}_A=\sum_j (\partial\hat{U}_A/\partial\theta_j)\dot\theta_j$ together with the variational equations of motion \eqref{eqn:ODE}, one can derive the identity
\begin{equation}
    \left\|\hat{R}(t)\right\|_F^2=\mathrm{Tr}\left[\hat{H}^2(t)\right]-\sum_jf_j\dot{\theta}_j
\end{equation}
Therefore,
\begin{equation}
\label{eqn:local_error_rate}
    \epsilon(t)=\frac{1}{2\sqrt{D}}\sqrt{\mathrm{Tr}\left[\hat{H}^2(t)\right]-\sum_jf_j\dot{\theta}_j}
\end{equation}
In Sec. \ref{sec:scalability}, we have shown how to compute $f_j$ and $g_{jk}$ for exponentially large Hilbert spaces using the operator-manifold projection. Since the variational equations of motion are $g\dot\theta=f$, the velocities $\dot\theta_j$ are obtained directly from the ODE integration with no additional cost. The trace $\mathrm{Tr}[\hat{H}(t)^2]$ is also analytically computable for the models studied in this work (and more generally, for any Hamiltonian expressible as a linear combination of Pauli strings or other orthogonal operator bases). Thus, the local error rate $\epsilon(t)$, and hence the rigorous error bound on $\eta(t)$, is fully scalable and can be monitored in real time during the simulation.

\subsubsection{Computational workflow}
\label{sec:workflow}
The complete variational method is summarized in Algorithm \ref{alg:variational_floquet}. Given a periodic Hamiltonian and a chosen operator pool, the algorithm integrates the variational EOMs over one period, switching between the exact diagonalization and the projected operator evaluation depending on the Hilbert-space dimension. Once the final parameters are obtained, the Floquet Hamiltonian is extracted.

As detailed in Sec. \ref{sec:err_analysis}, the accuracy of the approximation can be monitored throughout the simulation via the local error rate $\epsilon(t)$ and the accumulated error bound $\mathrm{AEB}(t)=\int_0^t ds\,\epsilon(s)$. Both quantities are computable without constructing the full Hilbert space, making them practical even for large systems. For instance, the AEB can serve as a stopping criterion or a quality check, allowing the user to enforce a desired precision (e.g., $\mathrm{AEB}(T) \le 1\%$) without requiring an exact benchmark.

\begin{algorithm}[t]
\SetAlgoLined
\SetKwInOut{Input}{Input}
\SetKwInOut{Output}{Output}

\caption{Variational action principle for Floquet Hamiltonians}
\label{alg:variational_floquet}

\Input{
  Periodic Hamiltonian $\hat H(t)$ with period $T$; 
  operator pool $\{\hat{\mathcal O}_j\}_{j=1}^{M}$; 
  initial parameters $\boldsymbol{\theta}(0)=\mathbf 0$; 
  integration tolerance.
}
\Output{Variational Floquet Hamiltonian $\hat H_F^{\mathrm{var}}$.}

\BlankLine
Initialize $t \gets 0$.\;

\While{$t < T$}{
  Construct $\hat A(\boldsymbol{\theta}(t)) = \sum_{j=1}^{M} \theta_j(t)\hat{\mathcal O}_j$.\;

  \If{Exact evaluation (moderate Hilbert-space dimension)}{
    Diagonalize $\hat A(\boldsymbol{\theta}(t))$ to obtain eigenvalues $\{\lambda_n\}$ and eigenvectors $\{\ket{\phi_n}\}$.\;

    Compute the quantum geometric tensor
    \[
    \begin{aligned}
    g_{jk} &= \mathrm{Tr}\left[\hat{\mathcal O}_j \times (\mathcal{K}^g \circ \hat{\mathcal O}_k)\right], \\
    \mathcal{K}^g_{mn} &= \operatorname{sinc}^2(\Delta_{mn}), \quad 
    \Delta_{mn} = \frac{1}{2}(\lambda_m - \lambda_n).
    \end{aligned}
    \]

    Compute the generalized force
    \[
    \begin{aligned}
    f_j &= \operatorname{Tr}\left[ (\mathcal{K}^f \circ \hat{\mathcal O}_j) \, \hat H(t) \right], \\
    \mathcal{K}^f_{mn} &= i\,\frac{e^{-2i\Delta_{mn}} - 1}{2\Delta_{mn}},
    \end{aligned}
    \]
    where the kernel is evaluated with the limiting value $\mathcal{K}^f_{mn}=1$ when $\Delta_{mn}=0$.
  }
  \Else{
    Use precomputed structure constants $\alpha_{jk}^{\ell}$ and overlap matrix 
    $\Phi_{jk} = \operatorname{Tr}(\hat{\mathcal O}_j \hat{\mathcal O}_k)$.\;

    Compute $\chi_{\ell p} = \sum_j \theta_j \alpha_{jp}^{\ell}$.\;

    Diagonalize $\chi = V \Lambda V^{-1}$.\;

    Compute
    \[
    g = 2\Phi \, V \, \frac{\cosh(\Lambda)-I}{\Lambda^2} \, V^{-1},
    \]
    and
    \[
    f = \Phi \, V \, \frac{I - e^{-\Lambda}}{\Lambda} \, V^{-1} \, \mathbf h(t),
    \]
    where $\mathbf h(t)$ are the coefficients of $\hat H(t)$ in the operator basis.
  }

  Solve $g\,\dot{\boldsymbol\theta} = f$ for $\dot{\boldsymbol\theta}$ and advance the integration (e.g., RK45).\;
}
\Return{$\displaystyle \hat H_F^{\mathrm{var}} = \frac{1}{T} \sum_{j=1}^{M} \theta_j(T) \, \hat{\mathcal O}_j$}
\end{algorithm}

\subsection{Magnus expansion}
\label{sec:Magnus}
An important and widely used approach to Floquet Hamiltonians is the Magnus expansion \cite{effective_HF_1}. The time-evolution operator in the Magnus expansion reads 
\begin{equation}
    \hat{U}_M(t)=e^{-i\hat{\Omega}(t)},
\end{equation}
where $\hat{\Omega}(t)=\sum_{j=1}^{\infty}\hat{\Omega}_j(t)$. The first three terms are:
\begin{equation}
    \hat{\Omega}_1(t)=\int_0^t dt_1 \hat{H}(t_1),
\end{equation}
\begin{equation}
    \hat{\Omega}_2(t)=-\frac{i}{2}\int_0^t dt_1\int_0^{t_1} dt_2\left[\hat{H}(t_1),\hat{H}(t_2)\right],
\end{equation}
and
\begin{equation}
    \hat{\Omega}_3(t)=-\frac{1}{6}\int_0^t dt_1\int_0^{t_1}dt_2\int_0^{t_2}dt_3\hat{\zeta}(t_1,t_2,t_3),
\end{equation}
where
\begin{align}
\hat{\zeta}(t_1,t_2,t_3)
&= \left[\hat{H}(t_1),\left[\hat{H}(t_2),\hat{H}(t_3)\right]\right] \nonumber\\
&\quad + \left[\hat{H}(t_3),\left[\hat{H}(t_2),\hat{H}(t_1)\right]\right] .
\end{align}
A sufficient condition for the convergence of Magnus series on the interval $[0,t]$ is \cite{Casas2007-zj}
\begin{equation}
    \int_0^t\, ds\,||\hat{H}(s)||<\pi
\end{equation}
where $||.||$ stands for the spectral norm. In practice, one truncates the series to a finite order $M$: $\hat{\Omega}(t)\approx\sum_{j=1}^{M}\hat{\Omega}_j(t)$. Such approximation is valid in the weak drive regime or high frequency limits. In other regimes, higher order terms become important, but their evaluation quickly becomes cumbersome due to the proliferation of nested commutators and time integrals.

If $\ket{n(t)}$ is an eigenstate of $\hat{\Omega}(t)$ with eigenvalue $\omega_n(t)$, the local error rate (Eq.\eqref{eqn:local_error_rate}) can be expressed as
\begin{equation}
    \epsilon(t)=\frac{1}{2\sqrt{D}}\sqrt{\sum_{n,m}|\mathcal{D}_{nm}(t)|^2}
\end{equation}
where
\begin{align}
     \mathcal{D}_{nm}(t)=&e^{-i\Delta_{nm}^+(t)}\frac{\sin\Delta_{nm}^-(t)}{\Delta_{nm}^-(t)}\langle n(t)|\hat{\dot{\Omega}}(t)|m(t)\rangle\nonumber\\
     &\quad -e^{-i\omega_m(t)}\langle n(t)|\hat{H}(t)|m(t)\rangle,
\end{align}
with $\Delta_{nm}^\pm(t)=(\omega_n(t)\pm\omega_m(t))/2$. Substituting this expression into Eq. \eqref{eqn:central_inequality} yields the accumulated error bound for the truncated Magnus expansion.

Once $\hat{\Omega}(t)$ has been obtained, the Floquet Hamiltonian is given by
\begin{equation}
    \hat{H}_F=\frac{1}{T}\hat{\Omega}(T)
\end{equation}

\subsection{Variational resummation of the Magnus expansion}
\label{sec:var_resumm}

The Magnus expansion provides a systematic perturbative approximation to the time-evolution operator, but its truncation at finite order discards all higher-order nested commutators, which can become important in regimes such as strong driving. A natural improvement is to start with the operator manifold generated by the low-order Magnus terms, yet allow the coefficients of those operators to be determined variationally rather than by the fixed perturbative series.

Concretely, suppose the truncated Magnus expansion involves a set of time-independent Hermitian basis operators $\{\hat{\mathcal{O}}_j\}_{j=1}^{M}$ such that the Magnus propagator is
\begin{equation}
\hat{U}_M(t)=e^{-i\sum_j \alpha_j(t)\hat{\mathcal{O}}_j},
\end{equation}
where $\alpha_j(t)$ are fixed time-dependent coefficients obtained from the Magnus series. Using this operator pool, the variational ansatz is
\begin{equation}
\hat{U}_A(\boldsymbol{\theta}(t))=e^{-i\sum_j \theta_j(t)\hat{\mathcal{O}}_j},
\end{equation}
with $\theta_j(t)$ determined by solving the EOM \eqref{eqn:ODE}. In contrast to the predetermined Magnus coefficients, the variational parameters are not fixed analytically; instead, they evolve according to the action principle, dynamically adjusting to minimize the instantaneous residual.

Higher-order Magnus terms generate nested commutators that often produce contributions lying within the same operator manifold. Neglecting those contributions may lead to errors or even divergence of the series. The variational procedure outlined above therefore serves as a \textit{resummation} of the operator weights, yielding the optimal coefficients within the chosen manifold. Moreover, because the variational equations minimize the instantaneous residual of the Schr\"odinger equation (see Appendix \ref{app:optimum_var_vs_mag}), the variational method yields the optimal local direction of evolution. This property, together with the numerical results presented in Secs. \ref{sec:LMG} and \ref{sec:driven_Ising}, will demonstrate that the variational resummation outperforms the conventional Magnus expansion.

\subsection{Numerical diagonalization in Sambe space}
In this section, we present a complementary method that provides numerically exact benchmark results for periodically driven quantum systems whenever the Fourier-mode truncation can be converged. Such systems are governed by the time-dependent Schr\"odinger equation
\begin{equation}
\label{eqn:SE}
    \hat{H}(t)\psi_\alpha(t)=i\partial_t\psi_\alpha(t),
\end{equation}
which admits solutions of the Floquet form
\begin{equation}
    \psi_\alpha(t)=e^{-i\varepsilon_\alpha t}\phi_\alpha(t),
\end{equation}
where $\phi_\alpha(t+T)=\phi_\alpha(t)$ is the time-periodic Floquet mode and $\varepsilon_\alpha$ is the quasienergy. The latter is directly accessible experimentally, for instance through absorption spectra \cite{Wang2013-cq,Rechtsman2013-tg}.

Substituting the Floquet ansatz into the Schr\"odinger equation \eqref{eqn:SE} yields the eigenvalue problem
\begin{equation}
    \hat{\mathcal{H}}_F\phi_\alpha(t)=\varepsilon_\alpha\phi_\alpha(t),
\end{equation}
where
\begin{equation}
    \hat{\mathcal{H}}_F=\hat{H}(t)-i\partial_t
\end{equation}
is the quasienergy operator acting on the extended Hilbert space $\mathcal{H}\otimes L^2([0,T])$.

Expanding the Floquet mode in a Fourier basis,
\begin{equation}
    \phi_\alpha(t)=\sqrt{\frac{\omega}{2\pi}}\sum_{n=-\infty}^{\infty}u_n^{(\alpha)}e^{in\omega t},
\end{equation}
transforms the eigenvalue problem into an infinite-dimensional algebraic system,
\begin{equation}
\label{eq:exact_sambe}
    \sum_{\ell}\Big[\hat{H}^{(j-\ell)}+\ell\omega\delta_{j\ell}\Big]u_\ell^{(\alpha)}
    =
    \varepsilon_\alpha u_j^{(\alpha)},
\end{equation}
where $\hat{H}^{(m)}$ is the $m$-th Fourier component of the Hamiltonian, given by
\begin{equation}
    \hat{H}^{(m)}
    =
    \frac{1}{T}\int_0^Tdt\,e^{-im\omega t}\hat{H}(t),
\end{equation}
and $\omega=2\pi/T$ is the driving frequency.

In practice, the Fourier indices are truncated to a finite set $j,\ell=-M,\ldots,M$, resulting in a $D(2M+1)$-dimensional extended Hilbert space, where $D$ is the dimension of the physical Hilbert space. The required value of $M$ depends on the system and driving parameters; in regimes of strong driving or at low frequencies, many Fourier sectors may be required to achieve convergence, making exact diagonalization computationally prohibitive due to the rapidly growing matrix dimensions, for example in strongly driven spin chains. This computational bottleneck severely limits the applicability of the Sambe-space approach to many-body systems.

In contrast, our non-perturbative variational framework works directly in the physical Hilbert space, projecting the dynamics onto a much smaller operator manifold whose size can be systematically tailored to the problem at hand. This circumvents the need to diagonalize the large matrices arising in the extended Hilbert space while retaining high accuracy.

\section{Benchmark models and results}
\label{sec:applications}
In this section, we apply our variational approach to derive the effective Floquet Hamiltonian for three prototypical periodically driven systems, each highlighting different aspects of the method: the Rabi model (two‑level system, non‑perturbative effect), the driven Lipkin–Meshkov–Glick (LMG) model (collective spin, symmetry reduction), and the driven Ising chain (many‑body system, scalability and operator truncation). For each model we compare our results with exact diagonalization (Sambe space) and with the truncated Magnus expansion. 

\subsection{Periodically driven Rabi model}
\label{sec:rabi}
Our first model is the following driven two-level system \cite{Rabi_Floquet_Magnus}\\

\begin{equation}
\label{eqn:rabi_model}
    \hat{H}_{\rm R}(t)=\frac{\omega_0}{2}\sigma_z+\kappa\cos(\omega t)\sigma_x
\end{equation}

where $\omega_0$ is the level splitting in the undriven system, $\kappa$ is the driving strength, and $\omega$ is the frequency of the periodic drive. It is common to use the rotating frame in which we write the time evolution as $\hat{U}(t)=e^{-i\omega t\sigma_z/2}\hat{U}_{\rm rot}(t)$ \cite{Rabi_Floquet_Magnus}. The time-dependent Hamiltonian in this frame reads:
\begin{equation}
    H_{\text{rot}}(t) = \frac{\Delta}{2}\sigma_z + \kappa\cos^2(\omega t)\sigma_x - \frac{\kappa}{2}\sin(2\omega t)\sigma_y,
\end{equation}
where $\Delta=\omega_0-\omega$. Using this Hamiltonian, we find the Floquet Hamiltonian up to third order Magnus expansion to be
\begin{equation}
    H_F^{\rm (mag)} = \tilde{\Delta}\sigma_z 
+ \frac{\kappa}{2}\left[ 1 + \frac{\Delta}{2\omega} - \frac{\kappa^2}{16\omega^2} - \frac{\Delta^2}{4\omega^2} \right]\sigma_x
\end{equation}
where $\tilde{\Delta}=\frac{\Delta}{2} - \frac{\kappa^2}{8\omega}$. Here, $\kappa^2/8\omega$ is known as the Bloch–Siegert shift \cite{Rabi_Floquet_Magnus}. Adding $\kappa\sin(\omega t)\sigma_y$ to Eq. \eqref{eqn:rabi_model} renders the model exactly solvable via the rotating frame transformation $\hat{U}(t)=e^{-i(\omega t/2)\sigma_z}\hat{U}_{\rm rot}(t)$, yielding the static Floquet Hamiltonian $\hat{H}_{\rm rot}=\frac{\omega_0-\omega}{2}\sigma_z+\kappa\sigma_x$.   

Next, we would like to use our variational approach to obtain the exact Floquet Hamiltonian. We choose the following variational ansatz
\begin{equation}
\label{eqn:Rabi_op_ansatz}
    \hat{A}(\boldsymbol{\theta}(t))=\theta_0(t)I+\theta_x(t)\sigma_x+\theta_y(t)\sigma_y+\theta_z(t)\sigma_z
\end{equation}
Where the variational parameters $\theta_0(t)$, $\theta_x(t)$, $\theta_y(t)$, and $\theta_z(t)$ are obtained by solving Eq. \eqref{eqn:ODE} on the interval $[0,T]$ with the initial condition that all parameters to vanish at $t=0$. 

A useful simplification arises from the structure of the QGT and generalized force. From Eq. \eqref{eq:g_single_integral}, one finds that $g_{0k}=0$ for $k\neq 0$, decoupling the EOM for $\theta_0(t)$ from the remaining parameters. Additionally, Eq. \eqref{eq:f_doubel_sum} gives $f_0=0$, which implies $\dot{\theta}_0(t)=0$. Together with $\theta_0(0)=0$, this yields $\theta_0(t)=0$ for all $t$. Consequently, the dynamics reduce to a system of three coupled differential equations for $\theta_x(t)$, $\theta_y(t)$, and $\theta_z(t)$:
\begin{equation}
\label{eqn:Rabi_EOM}
    \sum_{k\in\{x,y,z\}} g_{jk}\,\dot{\theta}_k(t) = f_j, \qquad j\in\{x,y,z\}.
\end{equation}
Once the solution over one period is obtained, the approximate Floquet Hamiltonian is given by
\begin{equation}
    \hat{H}_F^{\mathrm{var}} = \frac{1}{T}\hat{A}(\boldsymbol{\theta}(T)) = \frac{1}{T}\sum_{j=x,y,z}\theta_j(T)\sigma_j,
\end{equation}
where the identity contribution vanishes identically.

\begin{figure}
    \centering
    \includegraphics[width=\linewidth]{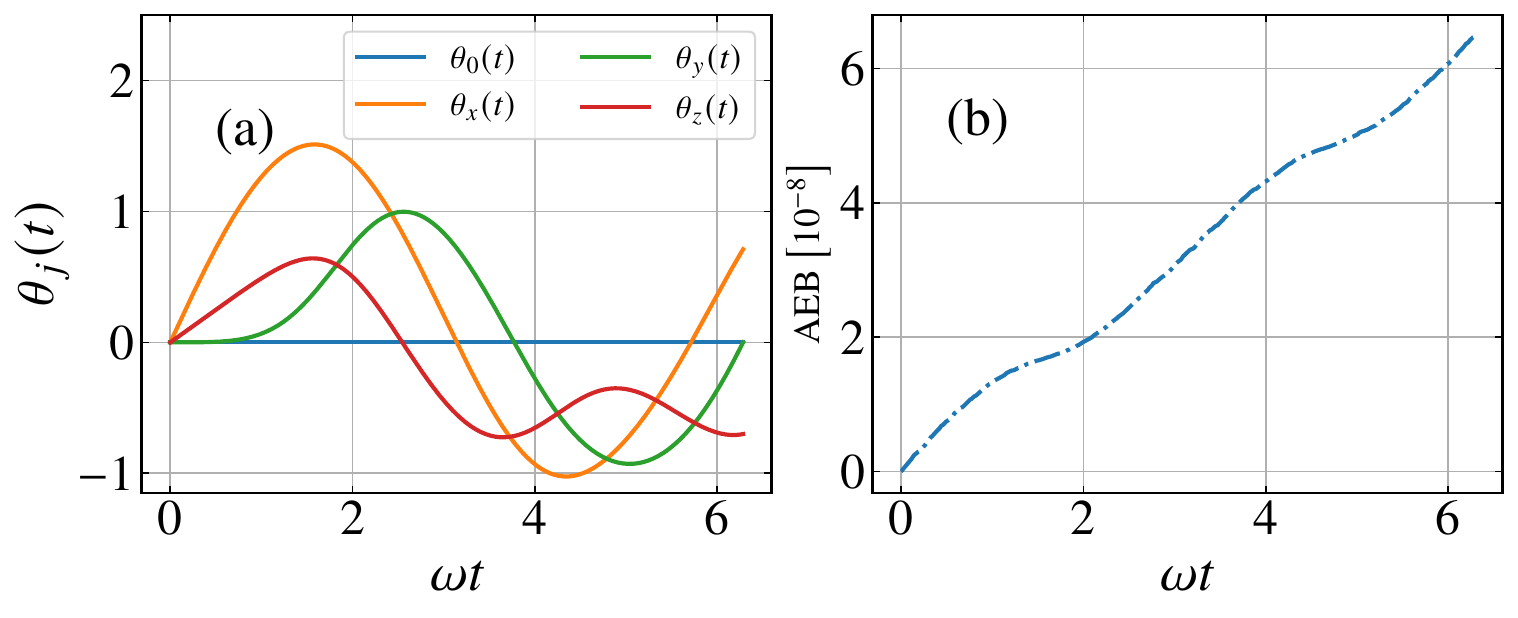}
    \caption{(a) Variational parameters for the Rabi model using the ansatz in Eq. \eqref{eqn:Rabi_op_ansatz}. Only $\theta_x(t)$ and $\theta_z(t)$ survive at the end of the period $T=2\pi/\omega$. (b) The accumulated error bound (AEB), defined as $\int_0^Tdt\,\epsilon(t)$, where $\epsilon(t)$ is the local error rate \eqref{eqn:local_error_rate}.  Parameters: $\kappa/\omega=1.5$, and $\omega_0/\omega=1$.}
    \label{fig:RabiVarparms}
\end{figure}

As an illustration, we obtain the variational parameters over one period for the case $\kappa/\omega=1.5$ and $\omega_0/\omega=1$, as shown in Fig. \ref{fig:RabiVarparms}(a). We observe that $\theta_0(T)$ and $\theta_y(T)$ vanish at the end of the period, while $\theta_x(T)$ and $\theta_z(T)$ are non-zero. This is true for arbitrary parameters as follows. Constructing the equations of motion for the Rabi model shows that the equation for $\theta_0(t)$ decouples from the rest of the system of differential equations and the initial condition $\theta_0(0)=0$ implies that $\theta_0(t)=0$ at all times. For $\theta_y(T)$, the symmetries of the Hamiltonian (reality and $\hat{H}(T-t)=\hat{H}(t)$) force the Floquet operator $\hat{U}_F(T)$ to be symmetric which eliminates the $\sigma_y$ terms (see Appendix \ref{app:rabi_symm}) and thus the Floquet Hamiltonian only includes $\sigma_x$ and $\sigma_z$ terms and takes the general form
\begin{equation}
    \hat{H}_F^{\rm (var)}=\frac{\theta_x(T)}{T}\sigma_x+\frac{\theta_z(T)}{T}\sigma_z
\end{equation}

Our numerical results also accurately capture the time-evolution operator over the entire period $t\in[0,T]$, as demonstrated by the plot of the accumulated error bound plot in Fig. \ref{fig:RabiVarparms}(b). The plot reveals that $\eta(t)$ has upper error bound of orders $10^{-8}$, indicating the high accuracy of the method. This is consistent with the closed $\mathfrak{su}(2)$ Lie algebra of the Rabi model: since the Baker–Campbell–Hausdorff series truncates, the exact evolution remains within the manifold, and our variational ansatz reproduces the exact dynamics to machine precision. 

\begin{figure}
    \centering
    \includegraphics[width=\linewidth]{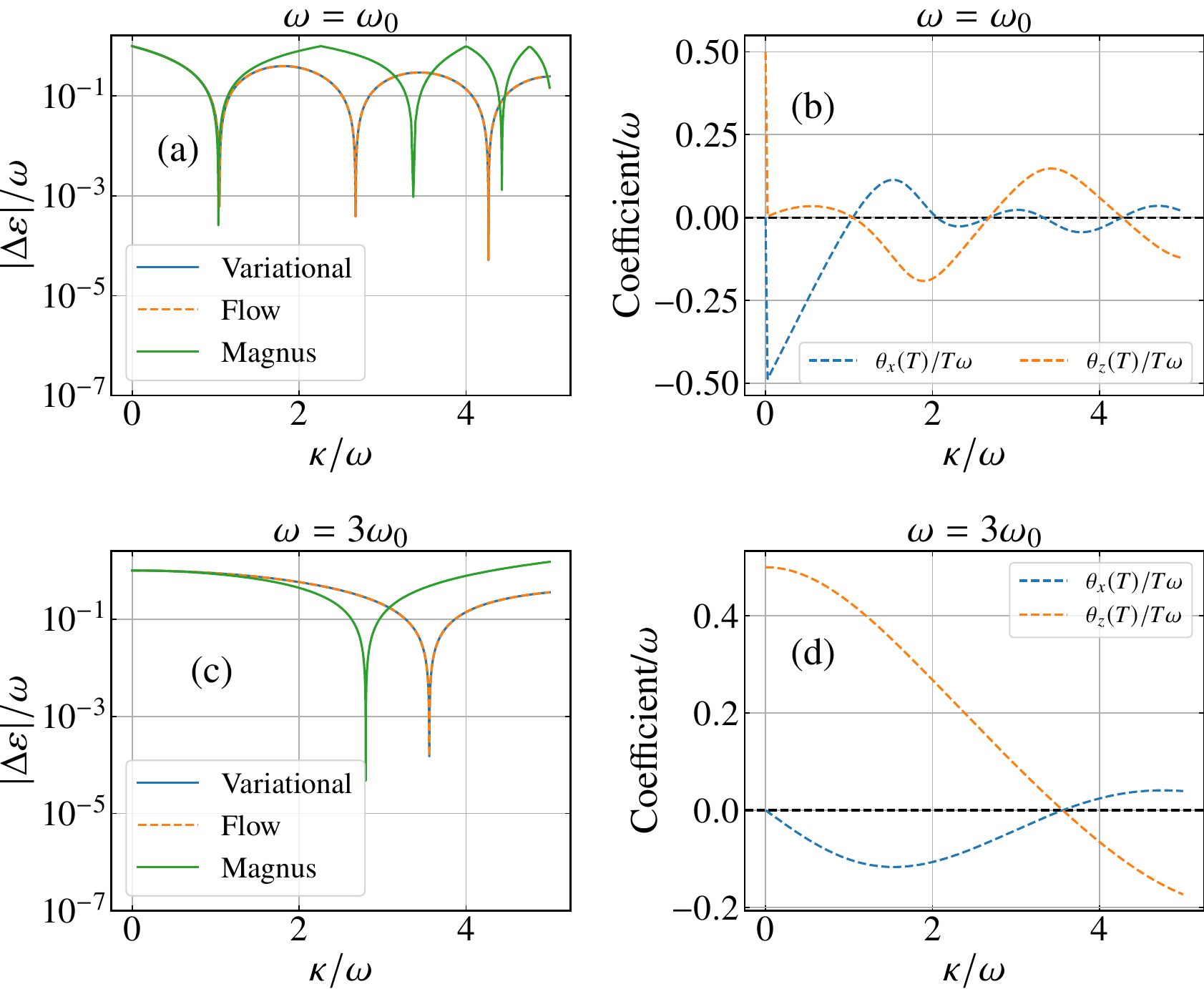}
    \caption{(a) A comparison of the quasi-energy gap (in units of $\omega$) for the Rabi model obtained via the variational method, the flow equation approach and the third order Magnus expansion as a function of $\kappa/\omega$ for the resonant case $\omega=\omega_0$ showing multiple peaks at specific values of the drive $\kappa/\omega$ where the gap vanishes. (b) The coefficients of the Floquet Hamiltonians obtained via the variational approach as functions of $\kappa/\omega$ for the resonant case. (c,d) same as (a,b) but with $\omega=3\omega_0$ (non-resonant regime).}
    \label{fig:Rabi_mag_vs_var_vs_flow}
\end{figure}

Next, we plot the quasi-energy gap for the Rabi model as a function of the drive strength in Fig. \ref{fig:Rabi_mag_vs_var_vs_flow}(a,c) for the resonant ($\omega=\omega_0$) and non-resonant ($\omega=3\omega_0$) cases obtained using the third order Magnus expansion, the variational approach, and the flow equation method (refer to Appendix \ref{app:flow_rabi} for further details) \cite{Michael_Flow_eq}. First, we observe that the three methods agree very well in the weak drive limit $\left({\rm small}\,\frac{\kappa}{\omega}\right)$ but then Magnus starts to break down at large drive ($\kappa/\omega>1$) as demonstrated in \ref{fig:Rabi_mag_vs_var_vs_flow}(a,c). This can be interpreted as the truncated series diverges in those limits and higher order terms in Magnus expansion become necessary. However, our method accurately captures the quasi-energies at large $\frac{\kappa}{\omega}$ as verified against the exact results obtained via the flow equation approach \cite{Michael_Flow_eq}. Furthermore, we have plotted the Floquet Hamiltonian parameters as functions of $\frac{\kappa}{\omega}$ as shown in Fig. \ref{fig:Rabi_mag_vs_var_vs_flow}(b,d). An important feature observed in Figs.~\ref{fig:Rabi_mag_vs_var_vs_flow}(a,c) is the appearance of sharp peaks at specific values of $\frac{\kappa}{\omega}$, indicating that the quasi-energy gap becomes zero at those points. Upon refining the grid around these $\frac{\kappa}{\omega}$ values, the peaks become increasingly sharper. Although we are limited by the numerical resolution, the trend clearly suggests that the gap vanishes exactly at certain \textit{critical} values of $\frac{\kappa}{\omega}$. This behavior is the well‑known phenomenon of coherent destruction of tunneling (CDT) \cite{CDT_phenomena,CDT_phenomena2,CDT_3}, where tunneling between the two states is completely suppressed.

\subsection{The driven Lipkin-Meshkov-Glick model}
\label{sec:LMG}
Our second example is the periodically driven Lipkin-Meshkov-Glick (LMG) model  \cite{Lipkin1965-mc,PRL_Krylov_Methods}
\begin{equation}
\hat{H}_{\rm LMG}(t) = -\frac{J}{2N}\sum_{i,j=1}^N\sigma_i^z\sigma_j^z - h\,\sin(\omega t)\sum_{j=1}^N\sigma_j^x,
\end{equation}
where $J$ is the long-range spin-spin interaction, $N$ is the number of particles or qubits, $h$ is the driving strength, and $\omega$ is the driving frequency. The above Hamiltonian can be rewritten in terms of the quasi-spin operators
\begin{equation}
    \hat{S}_z=\frac{1}{2}\sum_{j=1}^N\sigma_j^z, \quad \hat{S}_x=\frac{1}{2}\sum_{j=1}^N\sigma_j^x
\end{equation}
giving
\begin{equation}
\label{eq:LMG}
\hat{H}_{\rm LMG}(t) = -\frac{2J}{N}\hat{S}_z^2 - 2h\sin(\omega t)\,\hat{S}_x.
\end{equation}
which hugely reduces the Hilbert space dimension from $2^N$ to $N+1$.

An important symmetry of the above Hamiltonian is that it commutes with the spin-flip operator $\hat{P}=e^{i\pi\hat{S}_x}$ (i.e. $\mathbb{Z}_2$ symmetry) which flips the signs of $\hat{S}_y$ and $\hat{S}_z$. This implies that the time-evolution operator itself also commutes with $\hat{P}$ at all times. In our variational framework, if we choose the time-evolution ansatz as $\hat{U}_A(\boldsymbol{\theta}(t))=e^{-i\hat{A}(\boldsymbol{\theta}(t))}$, then we have $[\hat{A}(\boldsymbol{\theta}(t)),\hat{P}]=0$. This allows us to expand $\hat{A}(\boldsymbol{\theta}(t))$ in terms of a specific set of basis spin operators that commute with $\hat{P}$, providing a symmetry adaptive scheme for determining such a basis. This symmetry analysis reduces the computational cost as it avoids working with the full truncated manifold of operators (the truncated Lie Algebra). For example, if we restrict ourselves to spin operators up to quadratic order, then we have
\begin{equation}
\begin{aligned}
\hat{A}(\boldsymbol{\theta}(t))&=\theta_x(t)\hat{S}_x+\theta_{xx}(t)\hat{S}_x^2+\theta_{yy}(t)\hat{S}_y^2\nonumber\\
&+\theta_{zz}(t)\hat{S}_z^2+\theta_{yz}(t)\{\hat{S}_y,\hat{S}_z\}
\end{aligned}
\end{equation}
with the other linear ($\hat{S}_y$ and $\hat{S}_z$) and quadratic ($\{\hat{S}_x,\hat{S}_y\}$ and $\{\hat{S}_x,\hat{S}_z\}$) basis operators have been excluded by symmetry. To enlarge this basis one could include the cubic operators $\hat{S}_x^3$, $\{\hat{S}_x,\hat{S}_y^2\}$, $\{\hat{S}_x,\hat{S}_z^2\}$, and $\{\hat{S}_x,\{\hat{S_y},\hat{S}_z\}\}$. The corresponding variational parameters are $\theta_{xxx}(t)$, $\theta_{xyy}(t)$, $\theta_{xzz}(t)$, and $\theta_{xyz}(t)$, respectively. For a given ansatz $\hat{A}(t)$, we find the QGT $g_{jk}$ and the generalized force $f_j$ using Eqs. \eqref{eqn:gjk_final} and \eqref{eqn:fj_final}, respectively. Then, we plug in these results into Eq. \eqref{eqn:ODE} to integrate the equations of motion and obtain the variational parameters over one period $t\in[0,T]$. Using these results, we find corresponding Floquet Hamiltonian using $H^{\rm (var)}_F=\hat{A}(\boldsymbol{\theta}(T))/T$. 

Now, for the sake of comparison, we compute the lowest three contributions to Magnus expansion corresponding to our LMG model $\hat{\Omega}_1^{\rm LMG}(t)$, $\hat{\Omega}_2^{\rm LMG}(t)$, and $\hat{\Omega}_3^{\rm LMG}(t)$ given by
\begin{equation}
\label{eq:LMG_mag_1}
    \hat{\Omega}_1^{\rm LMG}(t)=-\frac{2Jt}{N}\hat{S}_z^2-\frac{2h}{\omega}(1-\cos(\omega t))\hat{S}_x,
\end{equation}

\begin{equation}
\label{eq:LMG_mag_2}
    \hat{\Omega}_2^{\rm LMG}(t)=\frac{2Jh}{N\omega}\left[t(1+\cos(\omega t))-\frac{2\sin(\omega t)}{\omega}\right]\{\hat{S}_y,\hat{S}_z\},
\end{equation}
and
\begin{align}
\label{eq:LMG_mag_3}
\hat{\Omega}_3^{\rm LMG}(t)
&= \frac{4J^2h}{3N^2}I_1(t)\left[\hat{S}_z^2,\left[\hat{S}_z^2,\hat{S}_x\right]\right] \nonumber\\
&\quad + \frac{4Jh^2}{3N}I_2(t)\left[\hat{S}_x,\left[\hat{S}_z^2,\hat{S}_x\right]\right].
\end{align}
The time dependence in $\hat{\Omega}_3^{\rm LMG}(t)$ is given by
\begin{equation}
    I_1(t)=\left(\frac{t^2}{2\omega}-\frac{6}{\omega^3}\right)(1-\cos(\omega t))+\frac{3 t \sin(\omega t)}{\omega^2},
\end{equation}
and
\begin{align}
I_2(t)
&= \frac{3}{\omega^3}\sin(\omega t)
   - \frac{2}{\omega^2}\, t \cos(\omega t)
   - \frac{9t}{4\omega^2} \nonumber\\
&\quad + \frac{3}{4\omega^3}\sin(2\omega t)
   - \frac{t}{4\omega^2}\cos(2\omega t).
\end{align}
It is important to verify that the above terms respect the symmetry imposed by the spin-flip operator described above. The first two contributions $\hat{\Omega}_1^{\rm LMG}(t)$ and $\hat{\Omega}_2^{\rm LMG}(t)$ expressed in terms of the basis operators discussed earlier and those automatically respect this symmetry. Now, for $\hat{\Omega}_3^{\rm LMG}(t)$, using spin commutation relations we find $\left[\hat{S}_z^2,\left[\hat{S}_z^2,\hat{S}_x\right]\right]=2\{\hat{S}_x,\hat{S}_z^2\}-\hat{S}_x$ and $\left[\hat{S}_x,\left[\hat{S}_z^2,\hat{S}_x\right]\right]=2\left(\hat{S}_y^2-\hat{S}_z^2\right)$. Thus, these commutators are expressed in terms of the known basis operators that commute with $\hat{P}$. Consequently, the low order Magnus expansion commutes with $\hat{P}$ as expected. The low order Floquet Hamiltonian obtained by Magnus expansion is
\begin{equation}
    \hat{H}^{\rm (Mag)}_F=\frac{\hat{\Omega}_1^{\rm LMG}(T)+\hat{\Omega}_2^{\rm LMG}(T)+\hat{\Omega}_3^{\rm LMG}(T)}{T}.
\end{equation}

\begin{figure}
    \centering
    \includegraphics[width=\linewidth]{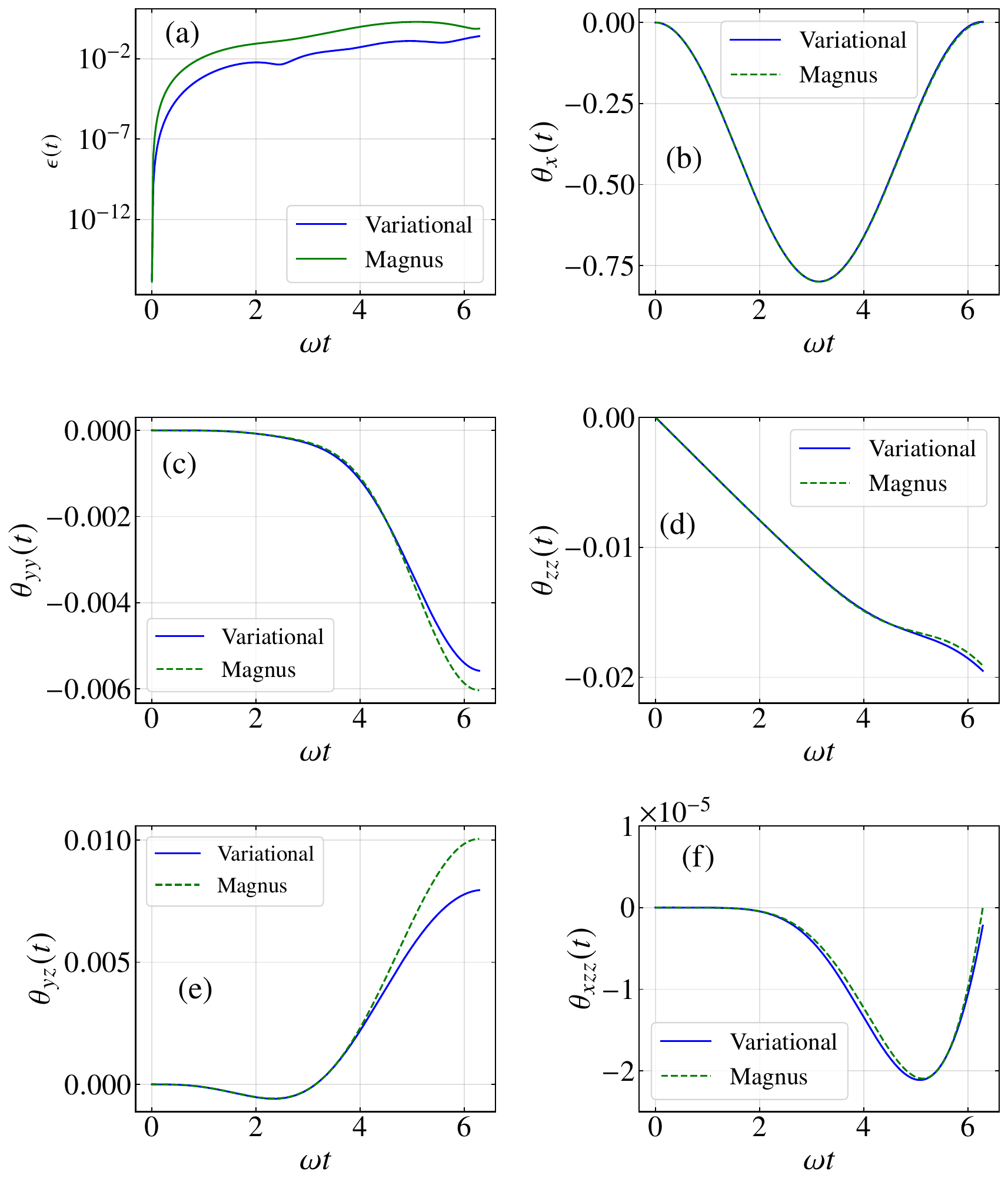}
    \caption{(a) The local error rate $\epsilon(s)$ \eqref{eqn:local_error_rate}, for the LMG model Eq. \eqref{eq:LMG} as a function of $\omega t$ for third-order Magnus expansion and the variational approach using the basis operators appearing in the truncated Magnus series. (b-f) Variational parameters as functions of $\omega t$ compared with the corresponding Magnus coefficients. Parameters: $J/\omega=h/\omega=0.2$, $N=100$.}    \label{fig:LMG_error_mag_vs_var_resum}
\end{figure}

As a first illustration, we select the operator pool $\{\hat{S}_x,\hat{S}_y^2,\hat{S}_z^2,\{\hat{S}_y,\hat{S}_z\},\{\hat{S}_x,\hat{S}_z^2\}\}$ for the variational ansatz. This set is motivated by the operators appearing in the third-order Magnus expansion of the LMG model. Figure \ref{fig:LMG_error_mag_vs_var_resum}(a) compares the local error rate of the variational approach with that of the Magnus expansion for parameters $J/\omega = h/\omega = 0.2$ and $N=100$. The variational method consistently yields a smaller error throughout the evolution, indicating that it tracks the exact generator of the dynamics more faithfully. This result substantiates the interpretation of the variational method as a resummation of the truncated Magnus series within the chosen operator manifold, as discussed in Sec.~\ref{sec:var_resumm}. 

Aside from the error metrics, we plot the variational parameters alongside the corresponding weights of the basis operators obtained from the Magnus expansion (Eqs. \eqref{eq:LMG_mag_1}-\eqref{eq:LMG_mag_3}) in Figs. \ref{fig:LMG_error_mag_vs_var_resum}(b-f). At early times, the variational parameters closely match the Magnus weights; as $\omega t$ increases, deviations emerge, notably in $\theta_{yy}(t)$ and $\theta_{yz}(t)$. This behavior is expected: the variational approach is designed to find the optimal parameters that make the time-evolution ansatz follow the actual Hamiltonian dynamics as closely as possible. Higher-order Magnus terms would contribute to the same set of operators, and the variational method captures these contributions without the need to compute complicated nested commutators.  

\begin{figure}
    \centering
    \includegraphics[width=\linewidth]{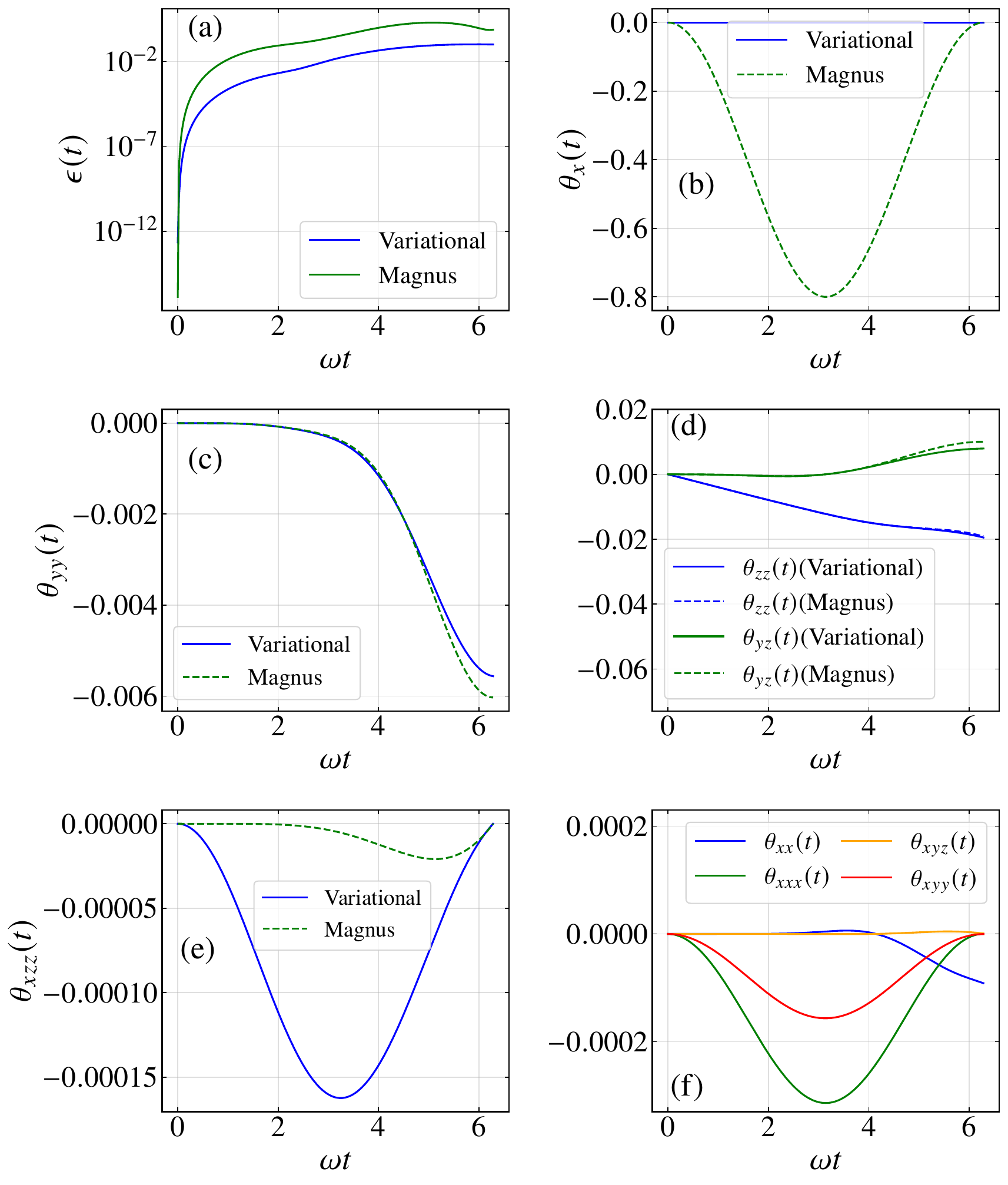}
    \caption{(a) A plot of the local error rate $\epsilon(s)$ \eqref{eqn:local_error_rate}, for the LMG model (Eq.\eqref{eq:LMG}) as a function of $\omega t$ for both third order Magnus expansion and the variational approach using the full pool basis operators up to cubic order in spin operators. (b,c,d,e,f) Plots of the variational parameters as functions of $\omega t$ and comparing them with the corresponding coefficients from Magnus series. Parameters: $J/\omega=h/\omega=0.2$, $N=100$.}
    \label{fig:LMG_error_mag_vs_var_fullCubic}
\end{figure}

Moving beyond the simplest ansatz, we enlarge the operator pool by including the missing quadratic operator $\hat{S}_x^2$, and the cubic operators $\hat{S}_x^3$, $\{\hat{S}_x,\hat{S}_y^2\}$, and $\{\hat{S}_x,\{\hat{S}_y,\hat{S}_z\}\}$. Figure~\ref{fig:LMG_error_mag_vs_var_fullCubic}(a) shows the local error rate for parameters $J/\omega = h/\omega = 0.2$ and $N=100$. As expected, this richer basis yields a consistently lower local error rate than the previous case shown in Fig.~\ref{fig:LMG_error_mag_vs_var_resum}(a). Although such extra operators would appear in higher-order Magnus expansions, the variational approach includes them without the cost of computing higher-order nested commutators and multi-dimensional time integrals. The variational parameters are obtained by solving the system of ordinary differential equations \eqref{eqn:ODE}, which ensures that the dynamics follows the Schrödinger equation as closely as possible.

We have also plotted the variational parameters in Figs. \ref{fig:LMG_error_mag_vs_var_fullCubic}(b-f). Interestingly, the variational approach finds that the weight of $\hat{S}_x$ is zero throughout the entire period $t\in[0,T]$ (Fig.~\ref{fig:LMG_error_mag_vs_var_fullCubic}(b)), whereas the Magnus expansion predicts a non-zero value for $t\in(0,T)$. This suggests that higher-order Magnus terms may contain cancellations not captured at third order. At the end of the period, however, both Magnus and the variational approach yield zero weight, indicating that $\hat{S}_x$ does not appear in the Floquet Hamiltonian $\hat{H}_F$ for this particular case. Similarly, in Fig. \ref{fig:LMG_error_mag_vs_var_fullCubic}(e), the variational solution yields larger contributions to $\{\hat{S}_x,\hat{S}_z^2\}$ than the Magnus result.

In Figs. \ref{fig:LMG_error_mag_vs_var_fullCubic}(c-d), we observe behaviour similar to that in Figs. \ref{fig:LMG_error_mag_vs_var_resum}(b) and (e-f): the variational and Magnus weights agree at short times, but deviations become apparent at larger $\omega t$. Furthermore, Fig. \ref{fig:LMG_error_mag_vs_var_fullCubic}(f) shows the weights of operators not present in the truncated Magnus series as functions of $\omega t$, revealing non-zero contributions. Thus, our variational approach extends beyond the Magnus expansion by allowing the inclusion of such operators, with their weights determined variationally through the action principle.

\begin{figure}
    \centering
    \includegraphics[width=\linewidth]{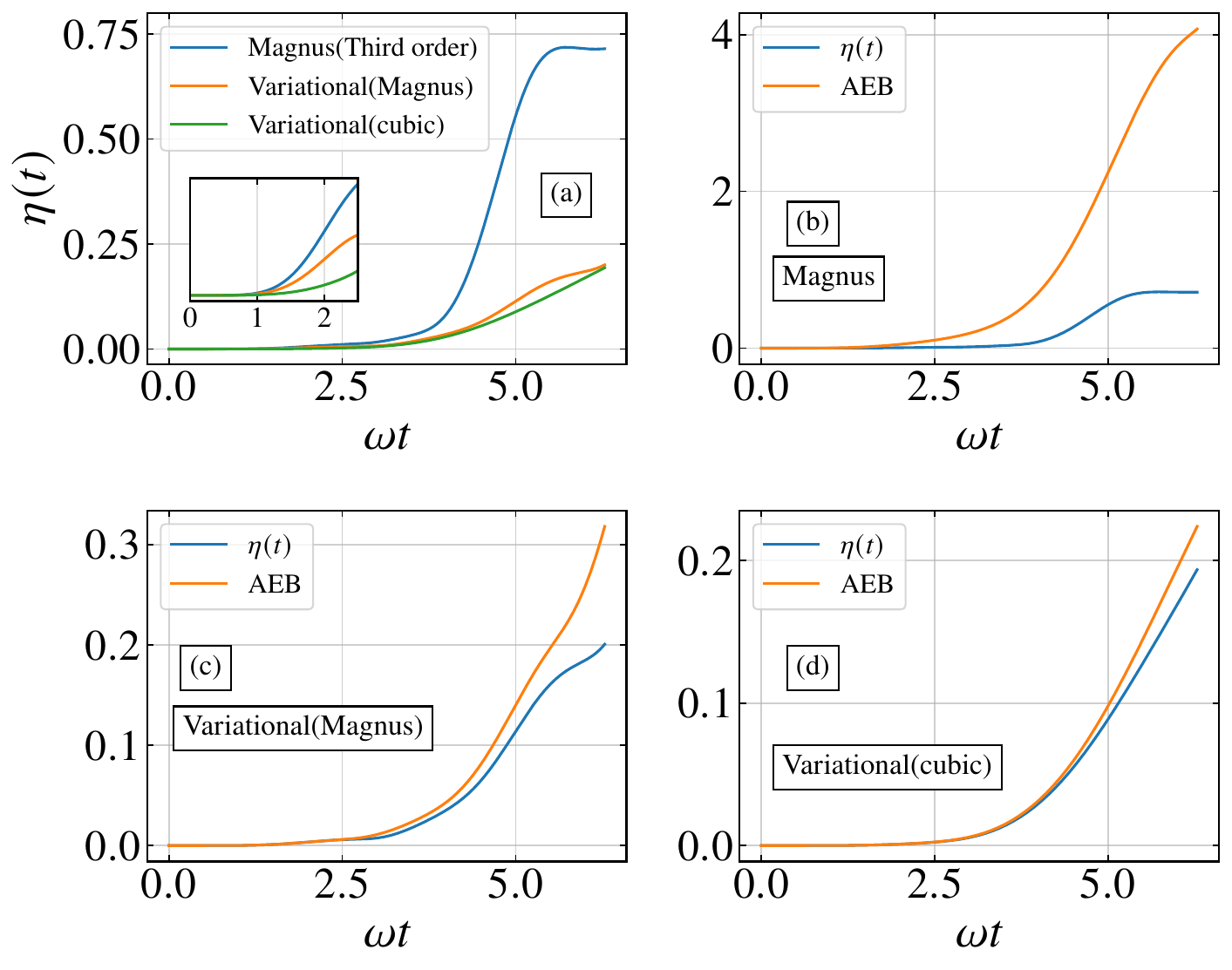}
    \caption{(a) Global error $\eta(t)$ defined in Eq. \eqref{eqn:global_error} as a function of $t$ for the LMG model \eqref{eq:LMG}. The results are shown for three time-evolution operator ansatz: (i) the third-order Magnus expansion, (ii) the variational approach using an operator pool generated from the Magnus expansion, and (iii) the variational approach using an operator pool generated from the Lie algebra of spin operators up to and including cubic order. (b-d) Plots of $\eta(t)$ versus the accumulated error bound (AEB) for the three time-evolution operator ansatz shown in (a), verifying the error bound given by Eq.~\eqref{eqn:central_inequality}. Parameters as $J/\omega=h/\omega=0.2$, and $N=100$.}
    \label{fig:LMG_global_errors_Jhp2_omega1_N100}
\end{figure}

Lastly, we plot the global errors for different time-evolution ansatz choices in Fig. \ref{fig:LMG_global_errors_Jhp2_omega1_N100} with parameters $J/\omega=h/\omega=0.2$ and $N=100$. The exact reference propagator $U_{\rm exact}(t)$ used to compute $\eta(t)$ is obtained by directly integrating the Schrödinger equation $\dot U=-iH(t)U$ with $U(0)=I$, which is numerically exact for the symmetric subspace dimension $N_s=N+1$.

In Fig. \ref{fig:LMG_global_errors_Jhp2_omega1_N100}(a), we compare $\eta(t)$ [Eq. \eqref{eqn:global_error}] for three approximations: (i) the third-order Magnus expansion, (ii) our variational approach using an operator pool constructed from the Magnus expansion, and (iii) the variational approach using a pool generated from the full Lie algebra of spin operators up to and including cubic order.

Remarkably, the variational ansatz using the Magnus-generated pool already yields a substantial reduction in the global error compared to the bare third-order Magnus expansion. This underscores that our variational procedure effectively resums the truncated Magnus series within the manifold, finding the trajectory that best approximates the true dynamics within the restricted operator space. A further improvement in accuracy is achieved when the pool is enlarged to include the full set of spin basis operators up to cubic order, confirming that systematically enriching the manifold drives the variational solution closer to the exact dynamics.  

In Figs. \ref{fig:LMG_global_errors_Jhp2_omega1_N100}(b–d), we plot the global error $\eta(t)$ against the accumulated error bound (AEB) for each of the three time-evolution ansatz shown in panel (a). The AEB is a time-dependent quantity derived analytically in Sec. \ref{sec:err_analysis} [Eq. \eqref{eqn:central_inequality}], and it provides a rigorous upper bound on the global error: $\eta(t) \le \text{AEB}$.

For all three approximations, third-order Magnus expansion, variational with Magnus-generated pool, and variational with full cubic spin algebra, the data points lie consistently below or on the AEB. This confirms that the analytically derived bound is satisfied throughout the entire time evolution. Notably, the variational ansatz yield significantly smaller errors than the Magnus expansion, while the bound remains tight across all cases, demonstrating that the inequality is both valid and practically useful for monitoring the quality of the approximation. The slight gap between $\eta(t)$ and the bound reflects the residual error not captured by the bound, consistent with the analysis in Sec. \ref{sec:err_analysis}.

\subsection{The driven Ising chain}
\label{sec:driven_Ising}

As a final and crucial benchmark, we consider the driven one-dimensional Ising model \cite{driven_Ising}, described by the Hamiltonian
\begin{equation}
\label{eqn:ising_model}
    \hat{H}_{\rm Ising}(t) = -J\sum_{j=1}^{N-1}\sigma_j^z\sigma_{j+1}^z - \frac{h}{2}\cos(\omega t)\sum_{j=1}^{N}\sigma_j^x,
\end{equation}
where $J$ is the Ising coupling, $h$ is the drive amplitude, and $\omega$ is the drive frequency. Unlike the Rabi and LMG models, which have Hilbert spaces of dimension $2$ and $N+1$, respectively, the Ising chain has an exponentially large Hilbert space of dimension $2^N$. This makes it an ideal testbed for assessing the scalability of our variational method and for demonstrating the operator-manifold projection technique developed in Sec.~\ref{sec:scalability}.

The Hamiltonian is invariant under the parity operator $\hat{P} = \bigotimes_{i=1}^{N} \sigma_i^x$, which allows us to restrict the basis operators to those that respect this symmetry, thereby reducing the size of the operator pool. For instance, if we restrict ourselves to a basis consisting only of single-body and nearest-neighbor two-body operators, parity symmetry dictates that only six operators lie in this manifold:
\begin{equation}
\label{eqn:quad_op_ising}
\begin{aligned}
\mathcal{O}_1 &= \sum_{j=1}^{N} \sigma_j^x, &
\mathcal{O}_2 &= \sum_{j=1}^{N-1} \sigma_j^x \sigma_{j+1}^x,\\
\mathcal{O}_3 &= \sum_{j=1}^{N-1} \sigma_j^y \sigma_{j+1}^y, &
\mathcal{O}_4 &= \sum_{j=1}^{N-1} \sigma_j^z \sigma_{j+1}^z,\\
\mathcal{O}_5 &= \sum_{j=1}^{N-1} \sigma_j^y \sigma_{j+1}^z, &
\mathcal{O}_6 &= \sum_{j=1}^{N-1} \sigma_j^z \sigma_{j+1}^y.
\end{aligned}
\end{equation}
Thus, we can employ the ansatz $\hat{A}(\boldsymbol{\theta}(t)) = \sum_{j=1}^6 \theta_j(t) \hat{\mathcal{O}}_j$, where the variational parameters $\theta_j(t)$ are real and determined by solving the equations of motion, Eq.~\eqref{eqn:ODE}, subject to the initial conditions $\theta_j(0)=0$ for all $j$.

An alternative approach for choosing the pool of operators is based on the truncated Magnus expansion. For the Ising model, the lowest two Magnus terms are
\begin{equation}
\label{eq:Ising_mag_1}
    \hat{\Omega}_1^{\rm Ising}(t) = -Jt\,\hat{\mathcal{O}}_4 - \frac{h\sin(\omega t)}{2\omega}\,\hat{\mathcal{O}}_1
\end{equation}
and
\begin{equation}
\label{eq:Ising_mag_2}
    \hat{\Omega}_2^{\rm Ising}(t) = -\frac{Jh}{2\omega^2}I(t)\bigl(\hat{\mathcal{O}}_5 + \hat{\mathcal{O}}_6\bigr),
\end{equation}
where $I(t) = \omega t\sin(\omega t) - 2(1-\cos(\omega t))$. The four operators appearing in these expressions can be used as a reduced basis for the variational approach, providing a direct link to the Magnus resummation discussed in Sec.~\ref{sec:var_resumm}.

Of course, one can enlarge the pool of operators by including additional nearest-neighbor two-body interactions or more complex multi-body interactions if deemed necessary. In particular, including three-body operators allows us to systematically improve the accuracy of the approximation. Using the short notation ${\rm XYZ} := \sum_{j=1}^{N-2} \sigma_j^x \sigma_{j+1}^y \sigma_{j+2}^z$, the 13 parity-invariant three-body nearest neighbor operators are
\begin{align}
\label{eqn:cubic_op_ising}
{\rm XXX},\, {\rm XZZ}, \,{\rm ZXZ},\,{\rm ZZX},\, {\rm XYY}, \,{\rm YXY},\,{\rm YYX}\nonumber\\
{\rm XYZ}, \,{\rm XZY},\,{\rm YXZ},\, {\rm YZX},\,{\rm ZXY},\, {\rm ZYX}.
\end{align}
Among these, only ${\rm ZXZ}$ appears in the third-order Magnus expansion, highlighting the flexibility of the operator manifold to extend beyond the perturbative Magnus operator content.

We now present numerical results for the driven Ising chain, focusing on two key aspects: (i) the improvement obtained by enlarging the operator pool from quadratic to cubic operators, and (ii) the demonstration of the resummation property by comparing variational results with truncated Magnus expansions.

\begin{figure}
    \centering
    \includegraphics[width=\linewidth]{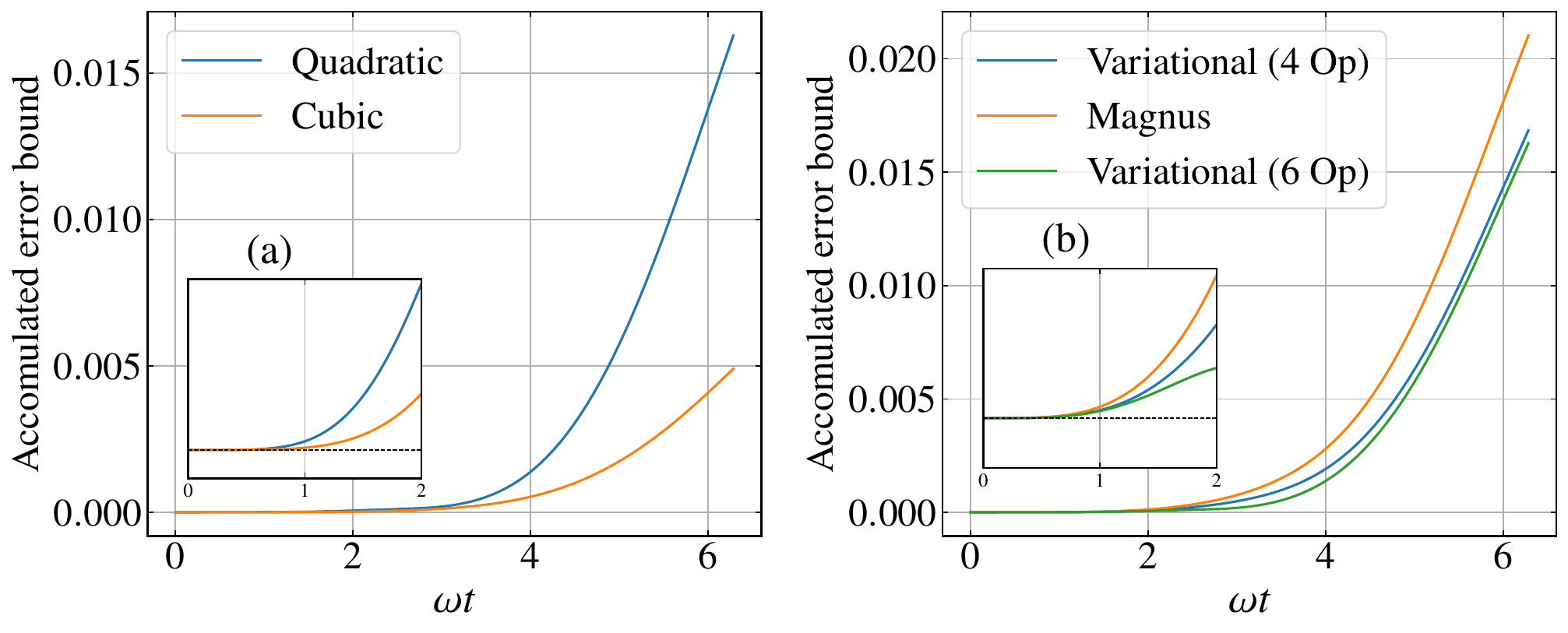}
    \caption{(a) A plot of the accumulated error bounds for the driven Ising model using the quadratic (two-body 1st NN interactions) and cubic (three-body interactions) basis operators. (b) Comparison between the accumulated error bounds obtained when using the truncated second order Magnus expansion and the variational approach results using the operator basis from Magnus expansion (4 Op) and using the allowed six two-body 1st NN interaction operators. Here we took the parameters as $J=1.0$, $h=0.5$, $\Omega=10$, and $N=8$.}
    \label{fig:Ising_combined_var_vs_mag_J1_hp5_omega10_N8}
\end{figure}

Figure~\ref{fig:Ising_combined_var_vs_mag_J1_hp5_omega10_N8} summarises the accumulated error bounds for $N=8$, $J=1.0$, $h=0.5$, and $\omega=10$. Panel (a) compares the error bounds obtained using the quadratic basis \eqref{eqn:quad_op_ising} and the cubic basis [\eqref{eqn:quad_op_ising} and \eqref{eqn:cubic_op_ising}]. The inclusion of three-body operators significantly tightens the error bound, confirming that the variational method systematically improves with an enlarged operator manifold. Panel (b) compares the variational method using the four operators from the Magnus series and the full six-operator quadratic basis against the second-order Magnus expansion. The variational approach with only four operators already outperforms Magnus, and the six-operator basis yields a further reduction in the error bound. This clearly demonstrates the resummation property discussed in Sec.~\ref{sec:var_resumm}: by allowing the coefficients of the Magnus operators to vary dynamically, the variational method effectively resums higher-order contributions that are missed by the truncated Magnus series.

\begin{figure}
    \centering
    \includegraphics[width=\linewidth]{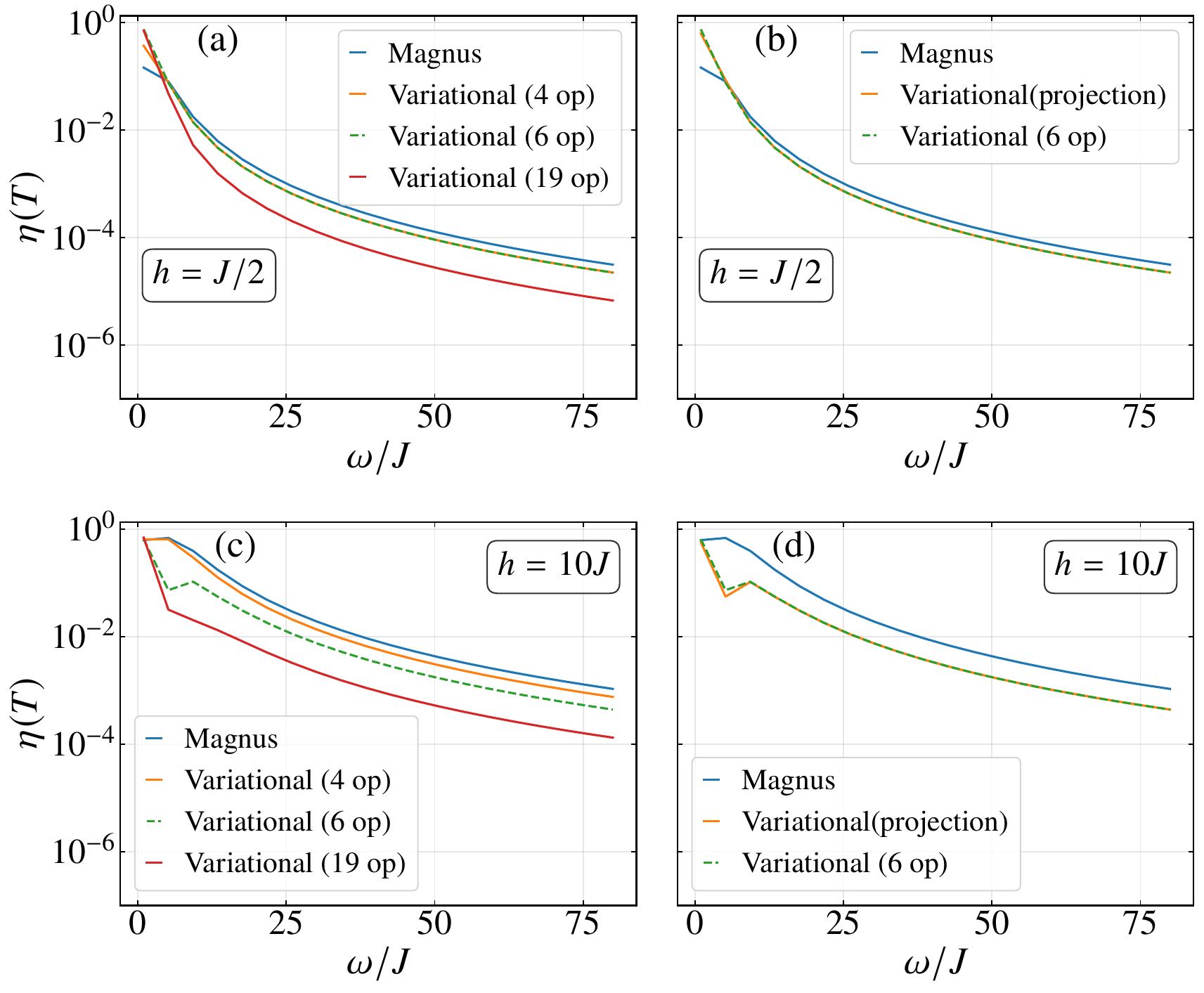}
    \caption{Global error $\eta(T)$ [Eq. \eqref{eqn:global_error}] of the approximate Floquet operator for the Ising model [Eq. \eqref{eqn:ising_model}] as a function of $\omega/J$. The left panels compare the second-order Magnus expansion with our variational approach using the four Magnus operators (4 op), the six basis operators [Eq. \eqref{eqn:quad_op_ising}], and the 19 basis operators [Eqs. \eqref{eqn:quad_op_ising} and \eqref{eqn:cubic_op_ising}]. The right panels compare the second-order Magnus expansion, the variational projection method of Sec.~\ref{sec:scalability} using the basis operators in Eq. \eqref{eqn:quad_op_ising}, and the variational approach using the same six basis operators. Panels (a,b) correspond to $h=J/2$, while panels (c,d) correspond to $h=10J$, with $N=5$.}
    \label{fig:Ising_operator_error}
\end{figure}

We now quantify the accuracy of our variational approach by computing the global error $\eta(T)$ [Eq.~\eqref{eqn:global_error}] in the Floquet operator, defined as the normalized Frobenius distance to the exact Floquet operator obtained by diagonalizing Eq. \eqref{eq:exact_sambe} with enough Fourier modes that ensures convergence. Figure~\ref{fig:Ising_operator_error}(a,c) presents $\eta(T)$ as a function of $\omega/J$ for the driven Ising model [Eq.~\eqref{eqn:ising_model}] with $N=5$, for a weak drive amplitude ($h=J/2$) and a strong drive amplitude ($h=10J$), respectively.

For the variational ansatz, we examine three progressively enriched operator pools: (i) the four operators appearing in the second order Magnus expansion; (ii) the six quadratic basis operators of Eq. \eqref{eqn:quad_op_ising}; and (iii) the full set of 19 operators combining quadratic and cubic terms from Eqs. \eqref{eqn:quad_op_ising} and \eqref{eqn:cubic_op_ising}. We compare these against the standard second order Magnus expansion.

At weak drive amplitude, $h=J/2$ [Fig.~\ref{fig:Ising_operator_error}(a)], all variational pools systematically outperform the Magnus expansion. Notably, the four- and six-operator pools yield nearly identical errors across the entire frequency range, suggesting that the dominant non-perturbative corrections are already captured by the Magnus-generated algebra. Enlarging the pool to 19 operators nonetheless reduces the error by roughly an order of magnitude, indicating that higher order commutators still measurably influence the Floquet dynamics.

The improvement persists and becomes even more pronounced at strong drive amplitude, $h=10J$ [Fig. \ref{fig:Ising_operator_error}(c)]. In this case, the four- and six-operator pools no longer coincide; the six quadratic operators yield better accuracy than the four Magnus operators. This highlights that the variational optimization can adaptively redistribute the effective Hamiltonian coefficients to capture drive-induced correlations beyond the standard Magnus truncation. As in the weak-drive case, the 19-operator pool consistently delivers the highest accuracy across all frequencies, confirming that augmenting the operator basis systematically reduces the variational error.

These results illustrate a central advantage of the proposed variational framework. Systematically enlarging the operator pool leads to a corresponding improvement in accuracy without requiring the explicit construction of higher-order Magnus expansions. Instead, increasingly higher-order contributions are incorporated through the enlarged operator manifold, providing a practical route to systematically improving the effective Hamiltonian in regimes where conventional high order Magnus expansions become prohibitively cumbersome.

In generating the results shown in Fig. \ref{fig:Ising_operator_error}(a,c), we leveraged the small system size to compute the generalized force $f$ and the quantum geometric tensor $g$ exactly, without any operator-space projection. This provides an ideal testbed for validating the manifold projection scheme introduced in Sec. \ref{sec:scalability}. To this end, we focus on the six-operator basis. We first derived analytical expressions for the overlap matrix $\Phi$ and the structure constants $\alpha_{jk}^\ell$ for arbitrary coupling parameters and system size $N$; the full details are provided in Appendix \ref{app:Ising_relations}. These closed-form expressions enable us to obtain the variational parameters for arbitrary $N$ using only $6 \times 6$ matrices. The same procedure can, of course, be applied to a larger operator pool.

In Fig. \ref{fig:Ising_operator_error}(b,d), we compare the projected variational results (using the six-operator manifold) against the exact variational results and the second-order Magnus expansion. Remarkably, for both $h=J/2$ and $h=10J$, the projected and exact variational curves are nearly indistinguishable and both outperform the Magnus expansion. This constitutes a crucial validation of our scalable framework: the projection onto the operator manifold faithfully reproduces the exact variational dynamics, even for strong drives and low frequencies. The method therefore holds considerable promise for applications in systems with large Hilbert space dimensions, where exact computations of $g$ and $f$ become infeasible.

\begin{figure}
    \centering
    \includegraphics[width=\linewidth]{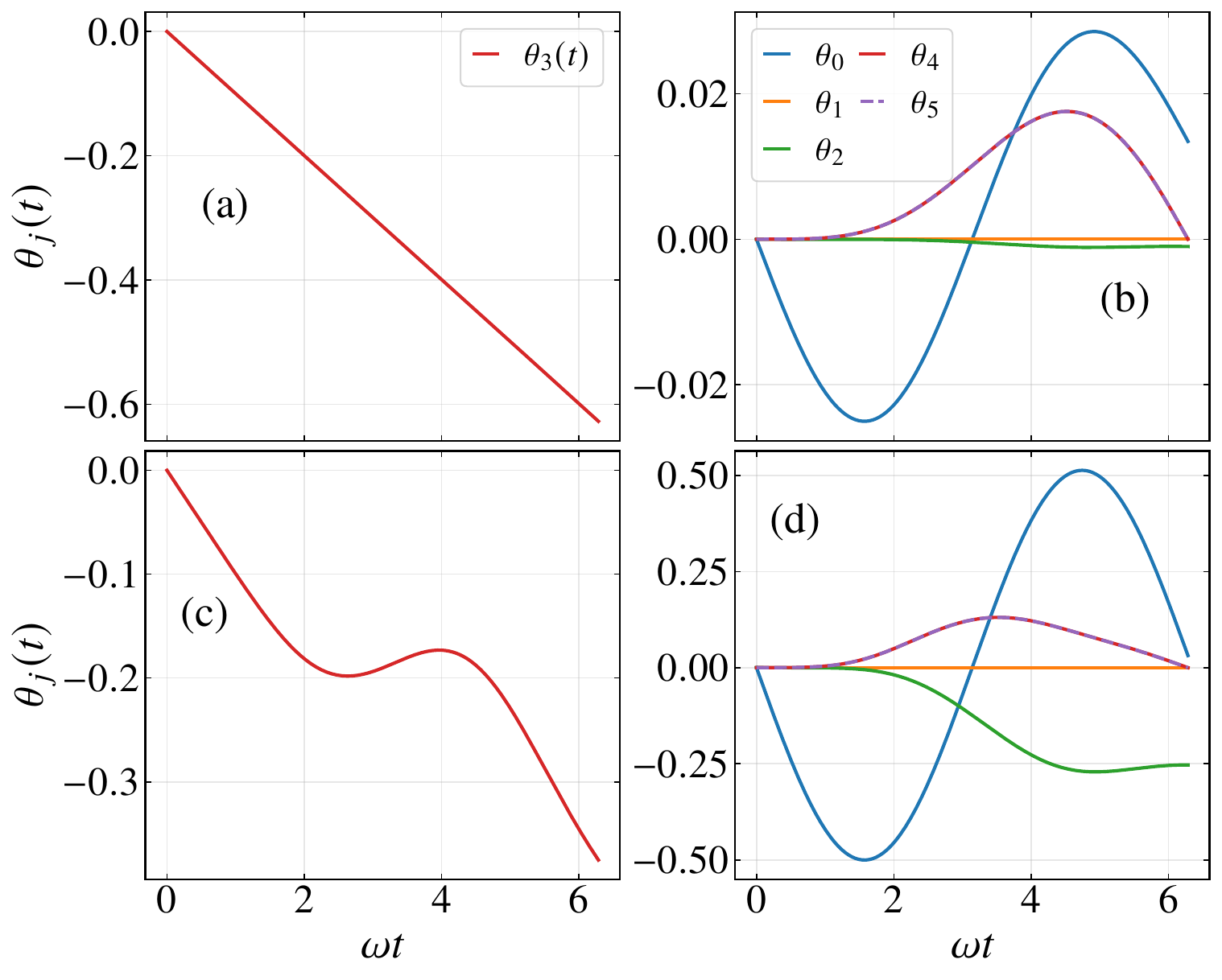}
    \caption{The variational parameters $\theta_j(t)$ for the six-operators basis [Eq. \eqref{eqn:quad_op_ising}], obtained using the manifold projection scheme of Sec. \ref{sec:scalability}. The system size is $N=100$, and the parameters are $J=1$, $\omega=10J$, with (a,b) $h=J/2$ and (c,d) $h=10J$.}
\label{fig:Ising_var_param_scalability}
\end{figure}

Finally, we demonstrate the scalability of our approach by computing the variational parameters for a chain of $N=100$ spins—a system size far beyond the reach of exact diagonalization. Fig. \ref{fig:Ising_var_param_scalability} shows the time-dependent coefficients $\theta_j(t)$ using the six-operator quadratic basis [Eq.~\eqref{eqn:quad_op_ising}]. The parameters are plotted for two representative drive amplitudes, $h=J/2$ (panels a,b) and $h=10J$ (panels c,d), with frequency fixed at $\omega=10J$.

Our results exhibit a well-defined thermodynamic limit, a consequence of the structure of the projected equations of motion. Because the overlap matrix $\Phi$ \eqref{eqn:phi_6op} cancels out of the equations of motion, $g\dot{\theta}=f$, the dynamics depend only on the structure constants $\alpha_{jk}^\ell$ \eqref{eqn:alphas_6op} and the Hamiltonian parameters. As shown in Appendix \ref{app:Ising_relations}, $\alpha_{jk}^\ell$ take values $-2$ and $2(1-1/N)$, with the $1/N$ corrections vanishing rapidly as $N$ grows. Consequently, $g$ and $f$ converge to $N$-independent constants already for modest chain lengths, and the variational parameters $\theta_j(t)$ become essentially independent of $N$ beyond some $N$. Consequently, the variational parameters in Fig. \ref{fig:Ising_var_param_scalability} already represent the thermodynamic limit.

\section{Conclusions}
\label{sec:conclusions}
In this work, we have developed a variational framework for approximating the time-evolution operator for generic time-dependent quantum systems. We have demonstrated its power by applying it to periodically driven systems, where it provides a non-perturbative route to constructing effective Floquet Hamiltonians. The method is based on an action principle for the propagator and uses a parameterized ansatz $\hat{U}_A(\boldsymbol{\theta}(t))$ built from a pool of physically motivated Hermitian operators. The variational parameters $\theta_j(t)$ are determined by solving a set of ordinary differential equations derived from the stationary action requirements. This approach is non‑perturbative, does not requires any constraints on the driving strength or frequency.

A key feature of the framework is its flexibility and systematic improvability. The operator manifold can be chosen based on symmetries, locality, or the specific physics of the problem. Enlarging this manifold (e.g., by including many‑body or longer-range operators) leads to a controlled improvement in the accuracy of the approximate propagator. Moreover, when the manifold is selected from the low‑order Magnus expansion, the variational method acts as a resummation of the Magnus series within that restricted operator space, significantly improving upon fixed order truncations. Unlike the Magnus expansion, which requires the evaluation of multi-dimensional time integrals of nested commutators to obtain the approximate propagator, our method allows us to include the expected operators explicitly and attach time-dependent weights that are determined variationally by solving a system of coupled differential equations.

The method also exhibits excellent scalability to many-body systems with exponentially large Hilbert spaces. By projecting the dynamics onto a fixed operator manifold of size $M$, all necessary quantities (the quantum metric $g_{jk}$, the generalized force $f_j$, and the equations of motion) can be evaluated using only $M\times M$ matrix operations. Thus, the explicit construction of the full Hilbert space is avoided, making the method applicable to large many‑body systems, such as driven spin chains with hundreds of sites, where exact diagonalization is not feasible.

We have illustrated the approach on three benchmark models: the two‑level Rabi model, the driven Lipkin–Meshkov–Glick model, and the driven one‑dimensional Ising chain. In all cases, the variational results show good agreement with exact diagonalization in Sambe space, where available, and outperform truncated Magnus expansions. The method correctly captures non‑perturbative phenomena such as coherent destruction of tunneling, and for the exactly solvable Rabi model, it reproduces the quasi-energy spectrum to machine precision. For the LMG and Ising models, the accuracy is quantitatively assessed via the global error $\eta(t)$ and the accumulated error bound.

While this manuscript was in preparation, we became aware of the recent work by Casares \textit{et al.} \cite{casares2026theorypracticetrotterproduct}, which introduces the Symmetry-Protected Randomized near-Integrable Trotter (SPRINT) formulas, a state-of-the-art framework for constructing optimized Trotter-Suzuki product formulas for static electronic-structure Hamiltonians in quantum chemistry. Although SPRINT is currently formulated for time-independent Hamiltonians, its underlying ideas including symmetry protection, randomization, and near-integrability, may also prove useful in extensions to time-dependent settings. The variational framework developed in the present work, which approximates the time-evolution operator within a physically motivated operator manifold, could naturally complement such approaches by enabling the dynamics to be projected onto reduced operator subspaces, thereby potentially lowering computational costs while mitigating errors associated with approximate time evolution. Exploring possible synergies between operator-manifold variational methods and optimized product formulas such as SPRINT represents a promising direction for future research.

Looking forward, this variational framework opens several promising directions. First, because the method does not rely on periodicity, it can be directly applied to non‑periodic time‑dependent Hamiltonians, offering a new tool for quantum control, adiabatic shortcuts, and open‑system simulations. Second, the scalable operator‑manifold formulation developed in Sec.\ref{sec:scalability} already enables efficient treatment of large many‑body systems; further improvements could involve adapting the basis to exploit additional symmetries or developing adaptive basis selection strategies guided by the accumulative error bound introduced in Sec. \ref{sec:err_analysis}. Finally, the resummation property demonstrated here suggests that the variational approach could be used to systematically improve other perturbative expansions in quantum dynamics.

In summary, we have formulated quantum dynamics as a variational problem in operator space, yielding a flexible, non-perturbative, and systematically improvable approach to time-dependent quantum systems. The extraction of effective Floquet Hamiltonians represents one important application, but the underlying formalism extends naturally to a wide class of driven quantum problems from non-periodic control protocols to large-scale many-body simulations.

\section{Acknowledgments}
IA and JPFL acknowledge the support of the Natural Sciences and Engineering Research Council of Canada (NSERC) RGPIN-2022-03882 and (NRC) AQC-200-1. MK acknowledges the support from the Applied Quantum Computing Challenge Program at the National Research Council of Canada.

\appendix
\begin{widetext}

\section{Derivations of $g$ and $f$ integral formulas}
\label{app:g_and_f_integrals}
To compute the derivative of the time-evolution ansatz $\hat{U}_A(t)=e^{-i\hat{A}(t)}$ with respect to the variational parameters, we use the following mathematical identity
\begin{equation}
    \frac{\partial}{\partial\lambda}e^{X(\lambda)}=\int_0^1 ds e^{(1-s)X(\lambda)}\frac{\partial X(\lambda)}{\partial \lambda}e^{sX(\lambda)}
\end{equation}
Using $X\to -iA$, we write
\begin{equation}
\label{eq:dU_dtheta}
    \frac{\partial}{\partial\theta_j}\hat{U}_A(t)=-i\int_0^1 ds e^{-i(1-s)\hat{A}(t)}\hat{\mathcal{O}}_je^{-is\hat{A}(t)}
\end{equation}
where we used $\hat{A}(t)=\sum_j \theta_j(t) \hat{\mathcal{O}}_j$ and thus $\partial \hat{A}(t)/\partial\theta_j=\hat{\mathcal{O}}_j$. Using the above identity into the QGT in Eq.\eqref{eqn:Qmetric}, we find
\begin{equation}
     g_{jk}=\int_0^1 ds'\int_0^1 ds\,\xi(s-s')
\end{equation}
with $\xi(s-s')=\mathrm{Tr}\left[\hat{\mathcal{O}}_je^{i(s-s')\hat{A}(t)}\hat{\mathcal{O}}_ke^{-i(s-s')\hat{A}(t)}\right]$, where we have used the cyclic property of the trace to simplify the expression of $\xi(s-s')$. Now, since the integrand depend on $s-s'$, we can do further simplifications by performing change of variables
\begin{equation}
    u=s-s', \quad v=s
\end{equation}
Thus, the Jacobian of the integrand becomes
\begin{equation}
    \mathcal{J}=\begin{bmatrix}
        \frac{\partial s}{\partial u} & \frac{\partial s}{\partial v}\\
        \frac{\partial s'}{\partial u} & \frac{\partial s'}{\partial v}
    \end{bmatrix}=\begin{bmatrix}
        0 & 1\\
        -1 & 1
    \end{bmatrix}
\end{equation}
Consequently, we have
\begin{equation}
    g_{jk}=\int du\int dv\,|{\rm det}(\mathcal{J})|\xi(u)=\int du\int dv\,\xi(u)
\end{equation}
It thus remains to determine the integration limits. Since $s,s'\in [0,1]$ and $u=s-s'$, then $u\in [-1,1]$. Furthermore, $s'=s-u$ and $0\leq s'\leq 1$, thus we have
\begin{equation}
    u\leq s\leq 1+u \to u\leq v\leq 1+u 
\end{equation}
But $v=s\in [0,1]$, so the interval for $v$ is $[0,1]\cap [u,1+u]$. To simplify further, let us take two cases. First, when $u\geq 0$, then $1+u\geq 1$ but $v$ cannot exceed $1$, thus the interval for $v$ becomes $[u,1]$ and hence $\int dv=1-u$. The second case is $u<0$, noting that $v\geq 0$, the interval for $v$ must be $[0,1+u]$, thus $\int dv=1+u$. Combing the two cases we have $\int dv=1-|u|$. Hence, the QGT becomes
\begin{equation}
\label{eq:single_integral_g}
    g_{jk}=\int_{-1}^{+1}du\,(1-|u|)\, \mathrm{Tr}\left[\hat{\mathcal{O}}_je^{iu\hat{A}(t)}\hat{\mathcal{O}}_ke^{-iu\hat{A}(t)}\right]
\end{equation}
Next, we want to find similar expression for the generalized force in Eq. \eqref{eqn:GF}. Using Eq. \eqref{eq:dU_dtheta}, we have
\begin{equation}
\label{eq:single_integral_f}
    f_j=\int_0^1 dz\,\mathrm{Tr}\left[e^{-iz\hat{A}(t)}\hat{\mathcal{O}}_je^{iz\hat{A}(t)}\hat{H}(t)\right].
\end{equation}
where we have made change of variable $z=1-s$.

\section{Minimization of the instantaneous residual by the variational method}
\label{app:optimum_var_vs_mag}
The exact time-evolution operator $\hat{U}(t)$ satisfies the Schr\"odinger equation $i\partial_t\hat{U}(t)=\hat{H}(t)\hat{U}(t)$. For a given time-evolution ansatz, we define the residual
\begin{equation}
    \hat{\mathcal{R}}=i\partial_t\hat{U}_A-\hat{H}(t)\hat{U}_A
\end{equation}
Defining the instantaneous squared residual
\begin{equation}
    \Phi=\frac{1}{2}||\hat{\mathcal{R}}||_F^2
\end{equation}
measures how accurate a given time-evolution ansatz is.  

By expressing the time-derivative of the variational ansatz as
\begin{equation}
    \hat{\dot{U}}_A=\sum_k\left(\frac{\partial\hat{U}_A}{\partial\theta_k}\right)\dot{\theta}_k
\end{equation}
where $\{\theta_j(t)\}$ is a set of variational parameters, and by plugging the above result into $\Phi$, we obtain
\begin{equation}
\Phi(\boldsymbol{\theta},\dot{\boldsymbol{\theta}})=\frac{1}{2}\sum_{j,k}g_{jk}\dot{\theta}_j\dot{\theta}_k-\sum_j {\rm Re}\{f_j\}\dot{\theta}_j+\frac{1}{2}\mathrm{Tr}\left[\hat{H}(t)^2\right]
\end{equation}
Now, let's find the stationary points of $\Phi$
\begin{equation}
    \frac{\partial\Phi}{\partial\dot{\theta_j}}={\rm Re}\left[\sum_k g_{jk}\dot{\theta}_k-f_j\right]=0
\end{equation}
Recalling that the EOM for the variational parameters is $\sum_k g_{jk}\dot{\theta}_k=f_j$, we conclude that the variational parameters are stationary points of $\Phi$. The second derivative of $\Phi$ is
\begin{equation}
    \frac{\partial^2\Phi}{\partial\dot{\theta}_j\partial\dot{\theta}_k}=g_{jk}
\end{equation}
It is sufficient to show that $g_{jk}$ is non-negative for $\Phi$ to be a convex function in velocities. Defining $\vec{\dot{\theta}}=(\dot{\theta}_1,\cdots,\dot{\theta}_n)^t$, we have
\begin{equation}
    \vec{\dot{\theta}}^T g\vec{\dot{\theta}}=\sum_{j,k}g_{j,k}\dot{\theta}_j\dot{\theta}_k=\left\|\sum_j \dot{\theta}_j\left(\frac{\partial\hat{U}_A}{\partial\theta_j}\right)\right\|_F^2\geq 0
\end{equation}
Thus $g$ is positive semi-definite, and $\Phi$ is a convex function of the velocities. Consequently, every stationary point of $\Phi$ is a global minimum with respect to the velocity variables $\dot{\boldsymbol{\theta}}$. Therefore, at each instant of time, the variational equations determine the velocity within the tangent space of the ansatz manifold that minimizes the instantaneous residual norm. In this sense, the method is locally optimal.

\section{Symmetry of the Floquet operator for the Rabi model}
\label{app:rabi_symm}
For the Rabi model Eq. \eqref{eqn:rabi_model}, the Hamiltonian has two symmetries
\begin{equation}
    \hat{H}_R(T-t)=\hat{H}_R(t), \quad \hat{H}_R^\top(t)=\hat{H}_R(t)
\end{equation}
where $\hat{H}^\top(t)$ is the transpose of $\hat{H}(t)$. Now, using Dyson's expansion for the time-evolution operator
\begin{equation}
    \hat{U}(T)=\hat{\mathcal{T}}e^{-i\int_0^T dt\,\hat{H}_R(t)}=I+\sum_{j=1}(-i)^j\int_0^T dt_1\int_0^{t_1}dt_2\cdots \int_0^{t_{j-1}} dt_j\hat{H}_R(t_1)\hat{H}_R(t_2)\cdots \hat{H}_R(t_j)
\end{equation}
where $0<t_j<\cdots<t_2<t_1<T$. Now, by calculating $\hat{U}_F(T)^\top$ and using the fact that $\hat{H}_R^\top(t)=\hat{H}_R(t)$, we get
\begin{equation}
    \hat{U}(T)^\top=I+\sum_{j=1}(-i)^j\int_0^T dt_1\int_0^{t_1}dt_2\cdots \int_0^{t_{j-1}} dt_j \hat{H}_R(t_j)\hat{H}_R(t_{j-1})\cdots \hat{H}_R(t_1)
\end{equation}
Using change of variables, 
\begin{equation}
    \tau_k=T-t_{j+1-k}
\end{equation}
and since $\hat{H}_R(T-\tau_k)=\hat{H}_R(\tau_k)$, we have
\begin{equation}
    \hat{H}_R(t_j)\hat{H}_R(t_{j-1})\cdots \hat{H}_R(t_1)=\hat{H}_R(\tau_1)\hat{H}_R(\tau_{2})\cdots \hat{H}_R(\tau_j)
\end{equation}
The region of integration becomes
\begin{equation}
    0<\tau_j<\cdots<\tau_2<\tau_1<T
\end{equation}
the transformation has a unit Jacobian, $|{\rm det}(\mathcal{J})|=1$. Hence, we obtain
\begin{equation}
    \hat{U}(T)^\top=I+\sum_{j=1}(-i)^j\int_0^T d\tau_1\int_0^{\tau_1}d\tau_2\cdots \int_0^{\tau_{j-1}} d\tau_j \hat{H}_R(\tau_1)\hat{H}_R(\tau_2)\cdots \hat{H}_R(\tau_j)=\hat{U}(T)
\end{equation}
For the Rabi model, the general time-evolution operator after one period is
\begin{equation}
    \hat{U}(T)=\alpha_0I+\alpha_x\sigma_x+\alpha_y\sigma_y+\alpha_z\sigma_z
\end{equation}
the requirement $\hat{U}(T)^\top=\hat{U}(T)$ implies that $\alpha_y=0$ since $\sigma_y^\top=-\sigma_y$. Consequently, the Floquet Hamiltonian will not have a $\sigma_y$ term.

\section{Floquet Hamiltonian for the Rabi model via the Flow equation approach}
\label{app:flow_rabi}

The flow equation approach is a non-perturbative method for obtaining Floquet Hamiltonians \cite{Michael_Flow_eq}. The central idea is to remove the time dependence of the Hamiltonian through an infinitesimal, step-by-step unitary transformation. This procedure yields a flow equation in terms of a flow parameter $s$ \cite{Michael_Flow_eq}:
\begin{equation}
\label{eqn:flow}
    \frac{d\hat{\mathcal{H}}(s,t)}{ds}=-\hat{V}(s,t)+i\int_0^t dt'\, [\hat{V}(s,t'),\hat{\mathcal{H}}(s,t)],
\end{equation}
where $\hat{\mathcal{H}}(s,t)$ is the ansatz Hamiltonian with $\hat{\mathcal{H}}(0,t)=\hat{H}(t)$, and
\begin{equation}
    \hat{V}(s,t)=\hat{\mathcal{H}}(s,t)-\frac{1}{T}\int_0^T dt\, \hat{\mathcal{H}}(s,t)
\end{equation}
is its time-dependent part. Constructing $\hat{\mathcal{H}}(s,t)$ typically requires an iterative, problem-specific identification of the operator manifold, as the ansatz must remain closed under the commutator generated by the flow. Once a closed manifold is found, the flow equation guarantees that $\hat{\mathcal{H}}(\infty,t)$ becomes independent of time, yielding the Floquet Hamiltonian $\hat{H}_F = \hat{\mathcal{H}}(\infty,t)$.

We now apply this method to the driven Rabi model described in Sec. \ref{sec:rabi}, with Hamiltonian $\hat H(t)=\frac{\omega_0}{2}\sigma_z + \kappa\cos(\omega t)\sigma_x$. We find that the ansatz $\hat{\mathcal{H}}(s,t)$ takes the form
\begin{equation}
\hat{\mathcal{H}}(s,t)= Z_0(s)\sigma_z+X_c(s)\cos(\omega t)\sigma_x+ Y_s(s)\sin(\omega t)\sigma_y+X_0(s)\sigma_x+Z_c(s)\cos(\omega t)\sigma_z
\end{equation}
where the $s$-dependent functions are obtained by substituting this ansatz into Eq. \eqref{eqn:flow}, yielding the coupled first-order differential equations
\begin{align}
\frac{dZ_0}{ds} &= \frac{2}{\omega} Y_s \left(X_0 - X_c\right), \label{eq:dZ0} \\[4pt]
\frac{dX_c}{ds} &= -X_c + \frac{2}{\omega} Y_s \left(Z_0 - Z_c\right), \label{eq:dXc} \\[4pt]
\frac{dY_s}{ds} &= -Y_s - \frac{2}{\omega} \left(X_0 Z_c - Z_0 X_c\right), \label{eq:dYs} \\[4pt]
\frac{dX_0}{ds} &= -\frac{2}{\omega} Y_s \left(Z_0 - Z_c\right), \label{eq:dX0} \\[4pt]
\frac{dZ_c}{ds} &= -Z_c - \frac{2}{\omega} Y_s \left(X_0 - X_c\right). \label{eq:dZc}
\end{align}
These are subject to the initial conditions
\begin{align}
    X_0(0)=0,\quad X_c(0)=\kappa,\quad Y_s(0)=0,\quad Z_0(0)=\omega_0/2,\quad Z_c(0)=0,
\end{align}
which follow directly from the requirement $\hat{\mathcal{H}}(0,t)=\hat{H}(t)$.

We solve this ODE system numerically as a function of $s$ and examine the limit $s\to\infty$. The flow equation guarantees that the coefficients attached to time-dependent terms vanish, i.e., $X_c(\infty)=Y_s(\infty)=Z_c(\infty)=0$. Hence, the Floquet Hamiltonian obtained via the flow equation approach reduces to
\begin{equation}
\label{eq:rabi_floquet_flow}
    \hat{H}_F^{(\rm Flow)}=Z_0(\infty)\sigma_z+X_0(\infty)\sigma_x.
\end{equation}
This effective Hamiltonian is used as a benchmark for our variational method in the Rabi model, as shown in Fig. \ref{fig:Rabi_mag_vs_var_vs_flow}. It is worth noting that for this simple Rabi model, the flow equation approach requires solving a system of five coupled differential equations (see above). By contrast, our variational method yields only three equations of motion as shown in Eq. \eqref{eqn:Rabi_EOM}. This structural reduction illustrates the lower computational cost and inherent efficiency of our approach.

\section{Evaluating $\alpha_{jk}^\ell$ and $\Phi$ for the driven Ising model}
\label{app:Ising_relations}
Considering the manifold spanned by the following six operators 
\begin{equation} 
\label{eqn:mf1_ising}
\begin{aligned}
\mathcal{O}_1 &= \sum_{j=1}^{N} \sigma_j^x, &
\mathcal{O}_2 &= \sum_{j=1}^{N-1} \sigma_j^x \sigma_{j+1}^x,\\
\mathcal{O}_3 &= \sum_{j=1}^{N-1} \sigma_j^y \sigma_{j+1}^y, &
\mathcal{O}_4 &= \sum_{j=1}^{N-1} \sigma_j^z \sigma_{j+1}^z,\\
\mathcal{O}_5 &= \sum_{j=1}^{N-1} \sigma_j^y \sigma_{j+1}^z, &
\mathcal{O}_6 &= \sum_{j=1}^{N-1} \sigma_j^z \sigma_{j+1}^y.
\end{aligned}
\end{equation}

We want first to obtain the structure constants $\alpha_{jk}^\ell$ defined through the commutator relation 
\begin{equation}
\left[\hat{\mathcal{O}}_j,\hat{\mathcal{O}}_k\right]=i\sum_\ell \alpha_{jk}^\ell \hat{\mathcal{O}}_\ell
\end{equation}

Starting with the identities  
\begin{equation}
\left[\sigma_j^x,\sigma_k^y\right]=2i\delta_{jk}\sigma_j^z,\quad\left[\sigma_j^x,\sigma_k^z\right]=-2i\delta_{jk}\sigma_j^y,\quad\left[\sigma_j^y,\sigma_k^z\right]=2i\delta_{jk}\sigma_j^x,\quad \left[\sigma_j^\alpha,\sigma_k^\alpha\right]=0
\end{equation}

We find that $[\hat{\mathcal{O}}_1,\hat{\mathcal{O}}_2]=0$, and the first non-zero commutator to be
\begin{equation}
\left[\hat{\mathcal{O}}_1,\hat{\mathcal{O}}_3\right]=\sum_{j=1}^{N}\sum_{k=1}^{N-1}\left[\sigma_j^x,\sigma_k^y\sigma_{k+1}^y\right]=\sum_{k=1}^{N-1}\left[\sigma_k^x,\sigma_k^y\right]\sigma_{k+1}^y+\sum_{k=1}^{N-1}\sigma_k^y\left[\sigma_{k+1}^x,\sigma_{k+1}^y\right]=2i\,(\hat{\mathcal{O}}_5+\hat{\mathcal{O}}_6)
\end{equation} 
which means $\alpha_{13}^5=\alpha_{13}^6=2$ and $\alpha_{13}^{\ell\neq 5,6}=0$. The rest of the commutators are
\begin{equation}
\label{eqn:Ising_commutators}
\begin{aligned}
[\hat{\mathcal{O}}_1,\hat{\mathcal{O}}_4] &= -2i\,\left(\hat{\mathcal{O}}_5+\hat{\mathcal{O}}_6\right),\\[2mm]
[\hat{\mathcal{O}}_1,\hat{\mathcal{O}}_5] &= 2i\,\left(\hat{\mathcal{O}}_4-\hat{\mathcal{O}}_3\right),\\[2mm]
[\hat{\mathcal{O}}_1,\hat{\mathcal{O}}_6] &= 2i\,\left(\hat{\mathcal{O}}_4-\hat{\mathcal{O}}_3\right),\\[2mm]
[\hat{\mathcal{O}}_2,\hat{\mathcal{O}}_3]& = 2i \sum_{k=1}^{N-2} \left( \sigma_k^x \sigma_{k+1}^z \sigma_{k+2}^y + \sigma_k^y \sigma_{k+1}^z \sigma_{k+2}^x \right),\\[2mm]
[\hat{\mathcal{O}}_2,\hat{\mathcal{O}}_4] &= -2i \sum_{k=1}^{N-2} \left( \sigma_k^x \sigma_{k+1}^y \sigma_{k+2}^z + \sigma_k^z \sigma_{k+1}^y \sigma_{k+2}^x \right),\\[2mm]
[\hat{\mathcal{O}}_2,\hat{\mathcal{O}}_5] &= 2i \sum_{k=1}^{N-2} \left( \sigma_k^x \sigma_{k+1}^z \sigma_{k+2}^z - \sigma_k^y \sigma_{k+1}^y \sigma_{k+2}^x \right),\\[2mm]
[\hat{\mathcal{O}}_2,\hat{\mathcal{O}}_6] &= 2i \sum_{k=1}^{N-2} \left( \sigma_k^z \sigma_{k+1}^z \sigma_{k+2}^x - \sigma_k^x \sigma_{k+1}^y \sigma_{k+2}^y \right),\\[2mm]
[\hat{\mathcal{O}}_3,\hat{\mathcal{O}}_4] &= 2i \sum_{k=1}^{N-2} \left( \sigma_k^z \sigma_{k+1}^x \sigma_{k+2}^y + \sigma_k^y \sigma_{k+1}^x \sigma_{k+2}^z \right),\\[2mm]
[\hat{\mathcal{O}}_3,\hat{\mathcal{O}}_5] &=2i\left( \sum_{j=2}^{N} \sigma_{j}^x + \sum_{j=1}^{N-2} \sigma_j^y \sigma_{j+1}^x \sigma_{j+2}^y \right),\\[2mm] 
[\hat{\mathcal{O}}_3,\hat{\mathcal{O}}_6] &=2i\left( \sum_{k=1}^{N-1} \sigma_k^x + \sum_{k=1}^{N-2} \sigma_k^y \sigma_{k+1}^x \sigma_{k+2}^y \right),\\[2mm]
[\hat{\mathcal{O}}_4,\hat{\mathcal{O}}_5] &=-2i\left( \sum_{k=1}^{N-1} \sigma_k^x + \sum_{k=1}^{N-2} \sigma_k^z \sigma_{k+1}^x \sigma_{k+2}^z \right),\\[2mm]
[\hat{\mathcal{O}}_4,\hat{\mathcal{O}}_6] &=-2i\left( \sum_{k=1}^{N-1} \sigma_{k+1}^x + \sum_{k=1}^{N-2} \sigma_k^z \sigma_{k+1}^x \sigma_{k+2}^z \right),\\[2mm]
[\hat{\mathcal{O}}_5,\hat{\mathcal{O}}_6] &= 0.
\end{aligned}
\end{equation}
By projecting the dynamics onto the manifold spanned by the six basis operators, we use the identity
\begin{equation}
\label{eqn:alphas_6op}
\alpha_{jk}^\ell=\frac{\mathrm{Tr}\left[[\hat{\mathcal{O}}_j,\hat{\mathcal{O}}_k]\hat{\mathcal{O}}_\ell\right]}{\mathrm{Tr}\left[\hat{\mathcal{O}}_\ell^2\right]}
\end{equation}
and find the non-zero $\alpha_{jk}^\ell$ to be 
\begin{equation}
\alpha_{14}^5=\alpha_{14}^6=-2,\quad \alpha_{15}^3=-\alpha_{15}^4=-2,\quad \alpha_{16}^3=-\alpha_{16}^4=-2,\quad \alpha_{35}^{1}=\alpha_{36}^{1}=-\alpha_{45}^{1}=-\alpha_{46}^{1}=2\frac{N-1}{N}
\end{equation}

It remains to compute the overlap matrix $\Phi_{jk}=\mathrm{Tr}\left[\hat{\mathcal{O}}_j\hat{\mathcal{O}}_k\right]$, which we find to be
\begin{equation}
\label{eqn:phi_6op}
\Phi = 2^N \begin{pmatrix}
N & 0 & 0 & 0 & 0 & 0 \\
0 & N-1 & 0 & 0 & 0 & 0 \\
0 & 0 & N-1 & 0 & 0 & 0 \\
0 & 0 & 0 & N-1 & 0 & 0 \\
0 & 0 & 0 & 0 & N-1 & 0 \\
0 & 0 & 0 & 0 & 0 & N-1
\end{pmatrix}
\end{equation}
The above results only needed to be computed once, then one can use the procedure introduced in Sec. \ref{sec:scalability} to obtain the corresponding variational parameters which are needed for extracting the Floquet Hamiltonian.  

\end{widetext}
\bibliography{ref.bib}
\end{document}